%% file: vtc0iopen.tex
\documentclass[leqno,draft,11pt]{article}
\usepackage{amssymb,amsmath,theorem,a4wide}

\def\sameenum{}

\input{defs}

\def\lh#1{\lVert#1\rVert}
\def\rcl#1{\tilde{#1}^\mathrm{real}}

\title{Open induction in a bounded arithmetic for~$\tc$}

\begin{document}
\maketitle
\begin{abstract}
The elementary arithmetic operations $+,\cdot,\le$ on integers are well-known to be computable in the weak complexity
class~$\tc$, and it is a basic question what properties of these operations can be proved using only $\tc$-computable
objects, i.e., in a theory of bounded arithmetic corresponding to~$\tc$. We will show that the theory $\vtc$ extended
with an axiom postulating the totality of iterated multiplication (which is computable in~$\tc$) proves induction for
quantifier-free formulas in the language $\p{+,\cdot,\le}$ ($\io$), and more generally, minimization for $\sig0$
formulas in the language of Buss's~$S_2$.
\end{abstract}

\section{Introduction}\label{sec:introduction}
Proof complexity is sometimes presented as the investigation of a three-way correspondence between propositional proof
systems, theories of bounded arithmetic, and computational complexity classes. In particular, we can associate to a
complexity class~$C$ satisfying suitable regularity conditions a theory~$T$ such that on the one hand, the provably
total computable functions of~$T$ of certain logical form define exactly the $C$-functions in the standard model of
arithmetic, and on the other hand, $T$ proves fundamental deductive principles such as induction and comprehension for
formulas that correspond to $C$-predicates. In this sense $T$ provides a formalization of $C$-feasible reasoning: we
can interpret provability in~$T$ as capturing the idea of what can be demonstrated when our reasoning capabilities are
restricted to manipulation of objects and concepts of complexity~$C$. The complexity class corresponding to a
``minimal'' theory that proves a given logical or combinatorial statement can be seen as a gauge of its proof
complexity. Then a particularly natural question is, given a function or predicate~$X$, which properties of~$X$ can be
proved by reasoning whose complexity does not exceed that of~$X$, that is, in a theory corresponding to the
complexity class for which $X$ is complete.

The main theme of this paper is what we can feasibly prove about the basic integer arithmetic operations
$+,\cdot,\le$. The matching complexity class is~$\tc$: $+$ and $\le$ are computable in $\Ac\sset\tc$, while $\cdot$
is in~$\tc$, and it is in fact $\tc$-complete under $\Ac$ (Turing) reductions. (In this paper, all circuit classes
like~$\tc$ are assumed $\LT$-uniform unless stated otherwise.) $\tc$ also includes many other functions
related to arithmetic. First, $+$ and $\cdot$ are also $\tc$-computable on rationals or Gaussian rationals.
An important result of Hesse, Allender, and Barrington~\cite{hab} based on earlier work by Beame
et al.~\cite{bch} and Chiu et al.~\cite{cdl} states that integer division and iterated
multiplication are $\tc$-computable. As a consequence, one can compute in~$\tc$ approximations of functions presented
by sufficiently nice power series, such as $\log$, $\sin$, or $x^{1/k}$, see e.g.\ Reif~\cite{reif}, Reif and
Tate~\cite{reif-tate}, Maciel and Th\'erien~\cite{mac-the:ser}, and Hesse et al.~\cite{hab}.

The more-or-less canonical arithmetical theory corresponding to~$\tc$ is $\vtc$ (see Cook and Nguyen~\cite{cook-ngu}).
This is a two-sorted theory in the setup of Zambella~\cite{zamb:notes}, extending the base $\Ac$-theory $V^0$ by an
axiom stating the existence of suitable counting functions, which gives it the power of~$\tc$. $\vtc$ is equivalent
($\rsuv$-isomorphic) to the one-sorted theory $\delt1\text-\M{CR}$ by Johannsen and Pollett~\cite{joh-pol:d1cr}, which
is in turn $\forall\exists\sig1$-conservative under the theory~$C^0_2$ \cite{joh-pol:c02}.

$\vtc$ can define addition and multiplication on binary integers, and it proves basic identities governing
these operations, specifically the axioms of discretely ordered rings (DOR). We are interested in what other properties of
integers expressible in the language $L_{\M{OR}}=\p{0,1,+,-,\cdot,\le}$ of ordered rings are provable in~$\vtc$, and in
particular, whether the theory can prove induction for a nontrivial class of formulas. Note that we should not expect
the theory to prove induction for bounded existential formulas, or even its weak algebraic consequences such as the
B\'ezout property: this would imply that integer gcd is computable in~$\tc$, while it is not even known to be
in~$\cxt{NC}$. However, this leaves the possibility that $\vtc$ could prove induction for \emph{open} (quantifier-free)
formulas of~$L_{\M{OR}}$, i.e., that it includes the theory~$\io$ introduced by Shepherdson~\cite{sheph}.

Using an algebraic characterization of open induction and a witnessing theorem for~$\vtc$, the provability of~$\io$ in
this theory is equivalent to the existence of $\tc$ algorithms for approximation of real or complex roots of
constant-degree univariate polynomials whose soundness can be proved in~$\vtc$. The existence of such algorithms in the
``real world'' is established in~\cite{ej:polyroot}, but the argument extensively relies on tools from complex analysis
(Cauchy integral formula, \dots) that are not available in bounded arithmetic, hence it is unsuitable for formalization
in~$\vtc$ or a similar theory.

The purpose of this paper is to demonstrate that $\io$ is in fact provable in a mild extension of~$\vtc$. The argument
naturally splits into two parts. We first formalize by a direct inductive proof a suitable version of the Lagrange
inversion formula (LIF), which was also the core ingredient in the algorithm in~\cite{ej:polyroot}. This allows us to
compute approximations of a root of a polynomial~$f$ by means of partial sums of a power series expressing the inverse function
of~$f$, but only for polynomials obeying certain restrictions on coefficients. The second part of the argument is model-theoretic, using
basic results from the theory of valued fields. The question whether a given DOR is a model
of~$\io$ can be reduced to the question whether the completion of its fraction field under a valuation induced by
its ordering is real-closed, and there is a simple criterion for recognizing real-closed valued fields. In our
situation, LIF ensures the relevant field is henselian, which implies that the criterion is satisfied.

We do not work with~$\vtc$ itself, but with its extension $\vtcim$ including an axiom ensuring the totality of iterated
multiplication. This theory corresponds to~$\tc$ just like~$\vtc$ does, as iterated multiplication is $\tc$-computable.
We need the extra axiom because it is not known whether $\vtc$ can formalize the $\tc$ algorithms for division and
iterated multiplication of Hesse et al.~\cite{hab}, and this subtle problem is rather tangential to the question of
open induction and root approximation. As explained in more detail in Section~\ref{sec:imul-div}, the $\imul$ axiom is
closely related to the integer division axiom~$\Div$ which is implied by~$\io$, hence its use is unavoidable in one way
or another. In terms of the original theory~$\vtc$, our results show that $\vtc\vdash\io$ if and only if
$\vtc\vdash\Div$.

We can strengthen the main result if we switch from~$L_{\M{OR}}$ to the language of Buss's one-sorted theories of
bounded arithmetic. By formalizing the description of bounded $\sig0$-definable sets due to Mantzivis~\cite{mantz},
$\vtcim$ can prove the $\rsuv$-translation of Buss's theory~$T^0_2$, and in fact, of the $\sig0$-minimization schema.
In other words, $T^0_2$ and $\sig0$-\Min\ are included in the theory $\delt1\text-\M{CR}+\imul$.

\section{Preliminaries}\label{sec:preliminaries}

A structure $\p{D,0,1,+,-,\cdot,\le}$ is an \emph{ordered ring} if $\p{D,0,1,+,-,\cdot}$ is a commutative (associative unital) ring, $\le$ is a linear
order on~$D$, and $x\le y$ implies $x+z\le y+z$ and $xz\le yz$ for all $x,y,z\in D$ such that $z\ge0$. If $D$ is an
ordered ring, $D^+$ denotes $\{a\in D:a>0\}$. A \emph{discretely ordered ring (DOR)} is an ordered ring~$D$ such that
$1$ is the least element of~$D^+$. Every DOR is an integral domain. An \emph{ordered field} is an ordered ring which is
a field. A \emph{real-closed field (RCF)} is an ordered field~$R$ satisfying any of the following equivalent conditions:
\begin{itemize}
\item Every $a\in R^+$ has a square root in~$R$, and every $f\in R[x]$ of odd degree has a root in~$R$.
\item $R$ has no proper algebraic ordered field extension.
\item The field $R(\sqrt{-1})$ is algebraically closed.
\item $R$ is elementarily equivalent to $\RR$.
\end{itemize}
(In a RCF, $\le$ is definable in terms of the ring structure, thus we can also call a field $\p{R,+,\cdot}$ real-closed
if it is the reduct of a RCF.) The \emph{real closure} of an ordered field~$F$ is a RCF $\rcl F\Sset F$ which is an
algebraic extension of~$F$. Every ordered field has a unique real closure up to a unique $F$-isomorphism.

The theory $\io$ consists of the axioms of ordered rings and the induction schema
\[\fii(0)\land\forall x\,(\fii(x)\to\fii(x+1))\to\forall x\ge0\,\fii(x)\]
for open formulas $\fii$ (possibly with parameters). An \emph{integer part} of an ordered field~$F$ is a discretely
ordered subring $D\sset F$ such that every element of~$F$ is within distance~$1$ from an element of~$D$.
The following well-known characterization is due to
Shepherdson~\cite{sheph}.
\begin{Thm}\th\label{thm:shep}
Models of~$\io$ are exactly the integer parts of real-closed fields.
\noproof\end{Thm}
The criterion is often stated with the real closure of the fraction
field of the model instead of a general real-closed field, but these
two formulations are clearly equivalent, as an integer part~$D$ of a
field~$R$ is also an integer part of any subfield $D\sset R'\sset R$.

In particular, models of~$\io$ are integer parts of their fraction
fields. This amounts to provability of the division axiom
\[\tag{$\Div$}\label{eq:div}
\forall x>0\,\forall y\,\exists q,r\,(y=qx+r\land0\le r<x)
\]
in~$\io$. (The uniqueness of $q$ and~$r$ holds in any DOR.)

We define $\Ac$ as the class of languages recognizable by a $\LT$-uniform family of polynomial-size
constant-depth circuits using~$\neg$ and unbounded fan-in $\land$ and $\lor$ gates, or equivalently, languages
computable by an $O(\log n)$-time alternating Turing machine with $O(1)$ alternations, or by a constant-time CRAM with
polynomially many processors~\cite{imm:cram}. If we represent an $n$-bit
binary string~$w$ by the finite structure $\cmb\p{{\{0,\dots,n-1\}},{<},{+},{\cdot},P_w}$, where $P_w(i)$ iff the $i$th bit
of~$w$ is~$1$, then $\Ac$ coincides with $\cxt{FO}$ (languages definable by first-order sentences). A language $B$ is
\emph{$\Ac$-reducible} to a language~$A$ if $B$ is computable by a $\LT$-uniform family of polynomial-size
constant-depth circuits using unbounded fan-in $\land$, $\lor$, $\neg$, and $A$-gates. The class of languages
$\Ac$-reducible to~$A$ is its \emph{$\Ac$-closure}.

$\tc$, originally introduced as a nonuniform class by Hajnal et al.~\cite{tc0}, is defined for our purposes as the
$\Ac$-closure of \task{Majority}. (Several problems $\tc$-complete under $\Ac$~reductions are noted in Chandra et
al.~\cite{ac0red}, any of these could be used in place of \task{Majority}.) Equivalently, $\tc$ coincides with languages
computable by $O(\log n)$-time threshold Turing machines with $O(1)$ thresholds, or by constant-time TRAM with
polynomially many processors \cite{par-sch}. In terms of descriptive complexity, a language is in~$\tc$ iff the
corresponding class of finite structures is definable in $\cxt{FOM}$, i.e., first-order logic with majority
quantifiers~\cite{founif}.

In connection with bounded arithmetic, it is convenient to consider not just the complexity of languages, but of predicates
$P(x_1,\dots,x_n,X_1,\dots,X_m)$ with several inputs, where $X_i$ are binary strings as usual, and $x_i$ are natural
numbers written in unary. It is straightforward to generalize $\Ac$, $\tc$, and similar classes to this context, see
\cite[\S IV.3]{cook-ngu} for details. Likewise, we can consider computability of functions: if $C$ is a complexity
class, a unary number function $f(\vec x,\vec X)$ is in~$FC$ if it is bounded by a polynomial in $\vec x$ and the
lengths of $\vec X$, and its graph $f(\vec x,\vec X)=y$ is in~$C$; a string function $F(\vec x,\vec X)$ is in~$FC$ if
the length of the output is polynomially bounded as above, and the bitgraph $G_F(\vec x,\vec X,y)\EQ(F(\vec x,\vec
X))_y=1$ is in~$C$. For simplicity, functions from $FC$ will also be called just $C$-functions.

We will work with two-sorted (second-order) theories of bounded arithmetic in the form introduced by
Zambella~\cite{zamb:notes} as a simplification of Buss~\cite{buss}. We refer the reader to Cook and
Nguyen~\cite{cook-ngu} for a general background on these theories as well as a detailed treatment of~$\vtc$, however,
we include the main definitions here in order to fix our notation.

The language $L_2=\p{0,S,+,\cdot,\le,\in,\lh\cdot}$ of second-order bounded arithmetic is a first-order language with
equality with two sorts of variables,
one for unary natural numbers, and one for finite sets thereof, which can also be interpreted as binary strings, or
binary integers. The standard convention is that variables of the first sort are written with lowercase letters
$x,y,z,\dots$, and variables of the second sort with uppercase letters $X,Y,Z,\dots$. While we adhere to this
convention in the introductory material on the theories and their basic properties, we will not follow it in the less
formal main part of the paper (we will mostly work with binary integers or rationals, and it looks awkward to write them all in
uppercase). The symbols $0,S,+,\cdot,\le$ of $L_2$ denote the usual arithmetic operations and relation on the unary
sort; $x\in X$ is the elementhood predicate, and the intended meaning of the $\lh X$ function is the least unary
number strictly greater than all elements of~$X$. This function is usually denoted as~$|X|$, however (apart from the
section on Buss's theories) we reserve the
latter symbol for the absolute value on binary integers and rationals, which we will use more often. We write $x<y$ as
an abbreviation for $x\le y\land x\ne y$.

Bounded quantifiers are introduced by
\begin{align*}
\exists x\le t\,\fii&\EQ\exists x\,(x\le t\land\fii),\\
\exists X\le t\,\fii&\EQ\exists X\,(\lh X\le t\land\fii),
\end{align*}
where $t$ is a term of unary sort not containing $x$ or~$X$ (resp.). Universal bounded quantifiers, as well as
variants of bounded quantifiers with strict inequalities, are defined in a similar way. A formula is $\Sig0$ if it
contains no second-order quantifiers, and all its first-order quantifiers are bounded. The $\Sig0$-definable predicates
in the standard model of arithmetic are exactly the $\Ac$ predicates.
A formula is $\Sig i$ if it
consists of $i$ alternating (possibly empty) blocks of bounded quantifiers, the first of which is existential, followed
by a $\Sig0$ formula. We define $\Pii i$ formulas dually. Similarly, a formula is $\Sigma^1_i$ ($\Pi^1_i$) if it
consists of $i$ alternating blocks of (possibly unbounded) quantifiers, the first of which is existential
(universal, resp.), followed by a $\Sig0$ formula\footnote{Notice that bounded second-order quantifiers still count
towards~$i$, so these formula classes do not correspond in the one-sorted setting to the usual arithmetical
hierarchy $\Sigma^0_i$, but to its restricted version where the formula after the main quantifier prefix is sharply
bounded. We follow~\cite{cook-ngu} in this usage; they only appear to define $\Sigma^1_1$, but we find it convenient to
extend this notation to higher levels as well.}.

The theory~$V^0$ in~$L_2$ can be axiomatized by the basic axioms
\begin{align*}
&x+0=x&&x+Sy=S(x+y)\\
&x\cdot0=0&&x\cdot Sy=x\cdot y+x\\
&Sy\le x\to y<x&&\lh X\ne0\to\exists x\,(x\in X\land\lh X=Sx)\\
&x\in X\to x<\lh X&&\forall x\,(x\in X\eq x\in Y)\to X=Y
\end{align*}
and the comprehension schema
\[\tag{$\fii$-\comp} \exists X\le x\,\forall u<x\,(u\in X\eq\fii(u))\]
for $\Sig0$ formulas~$\fii$, possibly with parameters not shown (but with no occurrence of~$X$). We denote the set~$X$
whose existence is postulated by $\fii$-\comp\ as $\{u<x:\fii(u)\}$. Using $\comp$, $V^0$ proves the induction and
minimization schemata
\begin{gather}
\tag{$\fii$-\ind} \fii(0)\land\forall x\,\bigl(\fii(x)\to\fii(x+1)\bigr)\to\forall x\,\fii(x),\\
\tag{$\fii$-$\Min$} \fii(x)\to\exists y\,\bigl(\fii(y)\land\forall z<y\,\neg\fii(z)\bigr)
\end{gather}
for $\Sig0$ formulas $\fii$. In particular, $V^0$ includes $\idz$ on the unary number sort.

Let $\p{x,y}$ be a $V^0$-definable pairing function on unary numbers, e.g., $\p{x,y}=(x+y)(x+y+1)/2+y$. We define
$X^{[u]}=\{x:\p{u,x}\in X\}$; this provides an encoding of sequences of sets by sets. We can encode sequences of unary
numbers by putting $X^{(u)}=\lh{X^{[u]}}$ (this is easily seen to be a $\Sig0$-definable function).
For convenience, we also extend the pairing function to (standard-length) $k$-tuples by
$\p{x_1,\dots,x_{k+1}}=\p{\p{x_1,\dots,x_k},x_{k+1}}$, and we write $X^{[u_1,\dots,u_k]}=X^{[\p{u_1,\dots,u_k}]}$,
$X^{(u_1,\dots,u_k)}=X^{(\p{u_1,\dots,u_k})}$.

$\vtc$ is the extension of~$V^0$ by the axiom
\[\forall n,X\,\exists Y\,\bigl(Y^{(0)}=0\land\forall i<n\,\bigl((i\notin X\to Y^{(i+1)}=Y^{(i)})
     \land(i\in X\to Y^{(i+1)}=Y^{(i)}+1)\bigr)\bigr),\]
whose meaning is that for every set~$X$ there is a sequence~$Y$ supplying the counting function
$Y^{(i)}=\card(X\cap\{0,\dots,i-1\})$.

Let $\Gamma$ be a class of formulas, and $T$ an extension of~$V^0$. A string function $F(\vec x,\vec X)$ is a
\emph{provably total $\Gamma$-definable function} of~$T$ if its graph is definable in~$\mathbb N$ by a formula
$\fii(\vec x,\vec X,Y)\in\Gamma$ such that $T\vdash\forall\vec x,\vec X\,\exists!Y\,\fii(\vec x,\vec X,Y)$; similarly
for number functions. If $\Gamma=\Sigma^1_1$, such functions are also called \emph{provably total recursive functions}
of~$T$. Note that one function may have many different definitions that are not $T$-provably equivalent; some of them
may be provably total, while other are not.

The provably total recursive functions of $V^0$ and~$\vtc$ are $\cxt F\Ac$ and $\cxt F\tc$, respectively. Moreover, we
can use these functions freely in the sense that if we expand the languages of the theories with the corresponding function
symbols, the resulting conservative extensions of $V^0$ and~$\vtc$ (respectively) prove the comprehension and induction
schemata for $\Sig0$ formulas of the expanded language; we will see more details in the next section.

Being $\Ac$, the ordering on binary integers is definable by a $\Sig0$ formula, and addition is provably
total in~$V^0$. Likewise, multiplication and iterated addition are provably total $\Sig1$-definable functions
of~$\vtc$. In fact, as shown in~\cite{cook-ngu}, the natural $\Sig0$ definitions of $X<Y$ and $X+Y$ provably satisfy
basic properties like commutativity and associativity in~$V^0$, and similarly, there are natural definitions of $X\cdot
Y$ and $\sum_{i<n}X^{[i]}$ provably total in~$\vtc$ such that $\vtc$ proves their basic properties, including the
inductive clauses
\begin{align*}
\sum_{i<0}X^{[i]}&=0,\\
\sum_{i<n+1}X^{[i]}&=\sum_{i<n}X^{[i]}+X^{[n]}.
\end{align*}
While Cook and Nguyen~\cite{cook-ngu} normally use second-sort objects to denote nonnegative integers, it will be
more convenient for us to make them represent all integers, which is easily accomplished by using one bit for sign. The
definitions of $<$, $+$, $\cdot$, and $\sum_{i<n}X^{[i]}$ can be adapted in a straightforward way to this setting so
that $\vtc$ still proves their relevant properties, that is, the axioms of discretely ordered rings.

\section{Iterated multiplication and division}\label{sec:imul-div}

As we already mentioned, it is not known whether $\vtc$ can formalize
the $\tc$ algorithms of Hesse, Allender, and Barrington~\cite{hab} for
integer division and iterated multiplication. In particular, it is not
known whether $\vtc$ proves the sentence~\ref{eq:div} (formulated for binary integers), which is a consequence
of~$\io$. This problem is rather tangential to the formalization of root
finding, whence we bypass it by strengthening our theory
appropriately.

It might seem
natural just to work in the theory~$\vtc+\Div$, however we will
instead consider an axiom stating the totality of iterated
multiplication in the following form:
\[\tag{$\imul$}\label{eq:imul}
\forall X,n\,\exists Y\,\forall i\le j<n\,
\bigl(Y^{[i,i]}=1\land Y^{[i,j+1]}=Y^{[i,j]}\cdot X^{[j]}\bigr).
\]
(The meaning is that for any sequence~$X$ of $n$~binary integers,
there is a triangular matrix~$Y$ with entries
$Y^{[i,j]}=\prod_{k=i}^{j-1}X^{[k]}$.) One reason is simply that we
need to use iterated multiplication at various places in the argument
(in particular, to compute partial sums of power series), and we do
not know whether $\vtc+\Div\vdash\imul$. The more subtle reason is
that we need the theory to be well-behaved in a certain technical
sense that we will describe in more detail below, and it turns out
that $\vtcim$ is the smallest well-behaved extension of $\vtc+\Div$.

Consider an extension $T\Sset V^0$ proving that a particular polynomially bounded recursive
(i.e., $\Sigma^1_1$-definable) function~$F$ is total, e.g.\  $\Div$ or~$\imul$.
While the most simplistic arguments employing~$F$ can get away with the mere fact that the value computed by~$F$ exists
for a particular input, usually we need more than that. For example, we may want to use induction on a
formula~$\fii(x)$ which involves $F$ applied to an argument depending on~$x$; since induction is obtained over~$V^0$ by
considering the least element of the set $\{x<a:\neg\fii(x)\}$, we effectively need comprehension for (simple enough)
formulas containing~$F$, say, $\Sig0(F)$-\comp.

From a computational viewpoint, it is desirable that we can combine provably total recursive functions in various
ways. For example, one of the basic $\tc$ functions is iterated addition, and a natural way how we would like to apply
it is to compute $\sum_{x<a}F(x)$ for a given provably total function~$F$. More generally, we want the class of provably
total recursive functions to be closed under $\Ac$ (or even $\tc$ in our case) reductions, and as a simple special
case, under \emph{parallel repetition}: if we can compute a function~$F(X)$, we want to be able to compute its
\emph{aggregate function} 
$F^*\colon\p{X_0,\dots,X_{n-1}}\mapsto\p{F(X_0),\dots,F(X_{n-1})}$ (where $n$ is a part of the input). In more logical terms, it
is desirable that $T$ is closed under the \emph{choice rule $\Sig0\text-\acr$}: if $T\vdash\forall X\,\exists Y\,\fii(X,Y)$,
where $\fii\in\Sig0$, then also $T\vdash\forall n\,\forall W\,\exists Z\,\forall i<n\,\fii(W^{[i]},Z^{[i]})$.
This is a derived rule corresponding to the axiom of choice, also called replacement or bounded collection:
\begin{equation}\tag{$\Sig0\text-\ac$}
\forall i<n\,\exists Y\le m\,\fii(i,Y,P)\to\exists Z\,\forall i<n\,\fii(i,Z^{[i]},P).
\end{equation}

Unfortunately, none of the desiderata mentioned in the last two paragraphs hold automatically, even for theories of the
simple form $V^0+\forall X\,\exists!Y\,F(X)=Y$ (note that $\vtc+\Div$ is of such form): this axiom implies the totality of functions
making a constant number of calls to~$F$, but we cannot a priori construct functions involving an unbounded number of
applications of~$F$, such as the aggregate function~$F^*$. However, Cook and
Nguyen~\cite{cook-ngu} show that the simple expedient of using~$F^*$ in the axiomatization instead of~$F$ leads to
theories satisfying all the properties above.
\begin{Def}\th\label{def:cn-th}
Let $\delta(X,Y)$ be a $\Sig0$-formula such that $V^0$ proves
\begin{gather*}
\delta(X,Y)\to\lh Y\le t(X),\\
\delta(X,Y)\land\delta(X,Y')\to Y=Y'
\end{gather*}
for some term $t(X)$. The \emph{Cook--Nguyen (CN) theory\footnote{In~\cite{cook-ngu}, $V(\delta)$ is
denoted $VC$, where the complexity class~$C$ is the $\Ac$-closure of~$F$, and it is called the minimal theory
associated with~$C$. We refrain from this terminology as the theory is not uniquely determined by the complexity class: it
depends on the choice of the $C$-complete function~$F$, and of a particular $\Sig0$-formula defining the graph
of~$F$ in~$\mathbb N$. In particular, both $\vtc$ and $\vtc+\imul$ are ``minimal'' theories for the same class ($\tc$), and it would be rather
confusing to call them as such.}} associated with~$\delta$ is
\[V(\delta)=V^0+\forall W,n\,\exists Z\,\forall i<n\,\delta(W^{[i]},Z^{[i]}).\]
(That is, if $F$ is a polynomially bounded function with an $\Ac$ graph defined by~$\delta$, which $V^0$ proves to be a
partial function, then $V(\delta)$ is axiomatized by the statement that the aggregate function~$F^*$ is total.)
\end{Def}

For example, $\vtc$ can be formulated as a CN theory, as shown in \cite[\S IX.3]{cook-ngu}.

\begin{Thm}\th\label{thm:cn-th}
Let $V(\delta)$ be a CN theory, and $F$ the function whose graph is defined by~$\delta$.
\begin{enumerate}
\item\label{item:6}
The provably total $\Sigma^1_1$-definable (or $\Sig1$-definable) functions of~$V(\delta)$ are exactly the functions in
the $\cxt{AC^0}$-closure of~$F$.
\item\label{item:7}
$V(\delta)$ has a universal definitional (and therefore conservative) extension~$\ob{V(\delta)}$ in a
language~$L_{\ob{V(\delta)}}$ consisting of $\Sig1$-definable functions of~$V(\delta)$. The theory $\ob{V(\delta)}$ has quantifier elimination for $\Sig0(L_{\ob{V(\delta)}})$-formulas, and it proves
$\Sig0(L_{\ob{V(\delta)}})$-\comp, $\Sig0(L_{\ob{V(\delta)}})$-\ind, and $\Sig0(L_{\ob{V(\delta)}})$-\Min.
\item\label{item:8} $V(\delta)$ is closed under $\Sig0$-\acr, and $V(\delta)+\Sig0\text-\ac$ is
$\Pi^1_2$-conservative over~$V(\delta)$. 
\end{enumerate}
\end{Thm}
\begin{Pf}
\ref{item:6} and~\ref{item:7} are Theorems IX.2.3, IX.2.14, and IX.2.16 in Cook and Nguyen~\cite{cook-ngu}.

\ref{item:8}: If $V(\delta)\vdash\forall X\,\exists Y\,\fii(X,Y)$ with~$\fii\in\Sig0$, there is an $L_{\ob{V(\delta)}}$-term $G(X)$ such
that $\ob{V(\delta)}\vdash\fii(X,G(X))$ by Herbrand's theorem, as $\ob{V(\delta)}$ is a universal theory, and $\fii$ is equivalent to an open formula. Then $\ob{V(\delta)}$, hence $V(\delta)$, proves
\[\forall W,n\,\exists Z\,Z=\{\p{i,y}:i<n,y\in G(W^{[i]})\}\]
using $\Sig0(L_{\ob{V(\delta)}})$-\comp.

The $\Pi^1_2$-conservativity of $\Sig0$-\ac\ over~$V(\delta)$ follows from the closure under $\Sig0$-\acr\ by cut
elimination. Alternatively, see~\cite[Thm.~4.19]{ej:vnc} for a model-theoretic proof generalizing
the result of Zambella~\cite{zamb:notes} for~$V^0$.
\end{Pf}

\begin{Lem}\th\label{lem:vtcim-cn}
\
\begin{enumerate}
\item\label{item:9} $\vtcim$ is a CN theory.
\item\label{item:10} $\vtcim\vdash\Div$. 
\end{enumerate}
\end{Lem}
\begin{Pf}
\ref{item:9}: The main observation is that $\vtcim$ proves the totality of the aggregate function of iterated
multiplication, that is,
\begin{equation}\tag{$\imul^*$}\label{eq:1}
\forall W,m,n\,\exists Z\,\forall k<m\,\forall i\le j<n\,
       \bigl(Z^{[k,i,i]}=1\land Z^{[k,i,j+1]}=Z^{[k,i,j]}\cdot W^{[k,j]}\bigr).
\end{equation}
Given $W,m,n$, put $X=\{\p{nk+j,x}:k<m,j<n,x\in W^{[k,j]}\}$ so that $X^{[nk+j]}=W^{[k,j]}$ for all $k<m$ and $j<n$,
and let $Y$ be as in~\eqref{eq:imul} for $X,mn$. Define
\[Z=\bigl\{\p{k,i,j,y}:k<m,i\le j\le n,y\in Y^{[nk+i,nk+j]}\bigr\},\]
so that $Z^{[k,i,j]}=Y^{[nk+i,nk+j]}$ for $k<m$ and $i\le j\le n$. Then $Z$ satisfies~\eqref{eq:1}.

Thus, $\vtcim=\vtc+\imul^*$. The latter looks almost like a CN theory, except that the graph of the
function specified in the axiom is not~$\Sig0$, as it involves multiplication. (The official definition also does not
allow an extra unary input, but this is benign as we could easily code $X,n$ into a single set.) There are several ways
how to get around this problem. For one, the whole machinery from \cite[\S IX.2]{cook-ngu} works fine if we
take~$\vtc$ instead of~$V^0$ as a base theory, and allow the use of $\Sig0(L_{\ob\vtc})$ formulas. Alternatively,
we can rewrite $\imul$ to incorporate the definition of multiplication, say
\begin{equation}
\tag{$\imul'$}\label{eq:8}
\begin{aligned}
\forall X,n\,\exists Y,Z\,\forall i\le j<n\,\forall x<\lh X\,
  \bigl(&Y^{[i,i]}=1\land Z^{[i,j,0]}=0\land Z^{[i,j,\lh X]}=Y^{[i,j+1]}\\
   &\land\bigl(x\notin X^{[j]}\to Z^{[i,j,x+1]}=Z^{[i,j,x]}\bigr)\\
   &\land\bigl(x\in X^{[j]}\to Z^{[i,j,x+1]}=Z^{[i,j,x]}+2^xY^{[i,j]}\bigr)\bigr),
\end{aligned}
\end{equation}
where $+$ and multiplication by~$2^x$ can be given easy $\Sig0$ definitions. Since the entries of~$Z$ can be
expressed as products of suitable $\Sig0$-definable sequences of integers, one can show in the same way as above that
$\imul'$, as well as the axiom $\imul'^*$ stating the totality of the corresponding aggregate function, is provable
in~$\vtcim$. Conversely, the CN theory $V^0+\imul'^*$ proves~$\vtc$ (as it implies the totality of usual
multiplication), hence it is equivalent to $\vtcim$.

\ref{item:10} can be shown by formalizing the reduction from~\cite{bch}. Assume that we want to find $\fl{Y/X}$, where
$X\ge1$. Choose $n,m>0$ such that $2^{n-1}\le X\le2^n$ and $Y\le2^m$, and put
\[Z=\sum_{i<m}(2^n-X)^i2^{n(m-1-i)}.\]
An easy manipulation of the sum shows that $XZ=2^{nm}-(2^n-X)^m$, hence
\[2^{nm}-2^{(n-1)m}\le XZ\le2^{nm}.\]
Put $Q=\fl{YZ/2^{nm}}$. Then
\[2^{nm}Y\ge XYZ\ge2^{nm}QX>XYZ-2^{nm}X\ge2^{nm}(Y-X-1),\]
hence $QX\le Y\le(Q+1)X$.
\end{Pf}

The more complicated converse reduction of iterated multiplication to division was formalized in bounded arithmetic by
Johannsen~\cite{joh:c02div} (building on Johannsen and Pollett~\cite{joh-pol:c02}), but in a different setting, so let us see what his result gives us here.
Johannsen works with a one-sorted theory $C^0_2[\M{div}]$, whose language consists of the usual Buss's
language for~$S_2$ expanded with $\dotminus$, $\M{MSP}$, and most importantly $\fl{x/y}$. It is axiomatized by a
suitable version of $\bas$, the defining axiom for division, the quantifier-free $\lind$ schema, and the axiom
of choice $\BB\sig0$ for $\sig0$ formulas in the expanded language.

We claim that $C^0_2[\M{div}]$ is $\rsuv$-isomorphic to the theory $\vtc+\M{DIV}+\Sig0\text-\ac$. We leave the
interpretation of the latter theory in~$C^0_2[\M{div}]$ to the reader as we will not need it, and focus on the
other direction. It is straightforward to translate the symbols of the language save division to the corresponding
operations on binary integers, and prove the translation of~$\bas$ in~$\vtc$. Of course, $\Div$ allows us to
translate the division function and prove its defining axiom, hence the only remaining problem is with the $\lind$ and $\BB$
schemata. Here we have to be a bit careful, as $\sig0$ (or even quantifier-free) formulas in the language of
$C^0_2[\M{div}]$ do not translate to $\Sig0$ formulas in the language of~$V^0$.

Let $\Div^*$ denote the axiom stating the totality of the aggregate function of division, or rather, of its expanded
version with witnesses for multiplication as in the proof of \th\ref{lem:vtcim-cn}, so that $T=\vtc+\Div^*$ is a CN
theory. By an application of choice, $\vtc+\Div+\Sig0\text-\ac$ proves $\Div^*$. Let $\ob T$ be the universal
conservative extension of~$T$ from \th\ref{thm:cn-th}, which includes function symbols for division and for $\tc$~functions
like multiplication. Since $\sig0$ formulas in the language of~$C^0_2[\M{div}]$ translate to $\Sig0(L_{\ob T})$
formulas, \th\ref{thm:cn-th} implies that $\ob T$, and therefore $T\sset\vtc+\Div+\Sig0\text-\ac$, proves the
translation of open (or even $\sig0$) $\lind$. As for the axiom of choice, every $\Sig0(L_{\ob T})$ formula is
equivalent to a $\Sig1$ formula in the language of~$V^0$, and $\Sig0$-\ac\ implies $\Sig1$-\ac, hence the translation
of $\BB\sig0$ is provable in $\ob T+\Sig0\text-\ac$, and thus in $\vtc+\Div+\Sig0\text-\ac$ by the conservativity of
$\ob T$ over~$T$.

This, together with provability of iterated multiplication in~$C^0_2[\M{div}]$, implies the
following:
\begin{Thm}[Johannsen~\cite{joh:c02div}]\th\label{thm:c02div}
$\vtc+\Div+\Sig0\text-\ac$ proves $\imul$.
\noproof\end{Thm}
\begin{Cor}\th\label{cor:div-cn-mul}
$\vtc+\imul=\vtc+\Div^*$ is the smallest CN theory including $\vtc+\Div$.
\end{Cor}
\begin{Pf}
Since $\vtc+\Div^*$ is a CN theory, \th\ref{thm:cn-th} implies that $\vtc+\Div+\Sig0\text-\ac$ is
$\Pi^1_2$-conservative over~$\vtc+\Div^*$, hence $\vtc+\Div^*\vdash\imul$ by \th\ref{thm:c02div}.
Conversely, every CN theory (such as $\vtc+\imul$, by \th\ref{lem:vtcim-cn}) that proves~$\Div$ also proves~$\Div^*$,
using its closure under $\Sig0$-\acr.
\end{Pf}
\begin{Cor}\th\label{cor:vtc-div-imul}
$\vtc\vdash\Div$ if and only if $\vtc\vdash\imul$.
\end{Cor}
\begin{Pf}
$\vtc$ is a CN theory.
\end{Pf}

The alert reader may have noticed that the reason why $\imul$ yields a CN theory while this is unclear for~$\Div$ is
not due to any deep property of iterated multiplication that would make it inherently better-behaved than division, but
because we made it so by formulating the axiom in the slightly redundant form using a triangular matrix of partial
products. There does not seem to be any particular reason we should expect to get a CN theory if we formulate the axiom
more economically, using only a one-dimensional array consisting of the products $\prod_{j<i}X^{[j]}$. In view of this,
the decision to axiomatize the theory using~$\imul$ rather than~$\Div^*$ is mostly a matter of esthetic preference and
convenience. Even in its triangular form, the $\imul$ axiom is a fairly natural rendering of the idea of computing
iterated products, whereas the usage of an aggregate function in~$\Div^*$ is overtly a technical crutch. Moreover, we
will be using iterated products more often than division, and while $\Div$ has a straightforward proof in $\vtc+\imul$
as indicated above, we would have to rely on the complicated argument from~\cite{joh:c02div} to derive~$\imul$ if we
based the theory on~$\Div^*$, making the main result of the paper less self-contained.

We mention another possibility for axiomatization of our theory, using the powering axiom
\[\tag{$\M{POW}$} \forall X,n\,\exists Y\,\forall i<n\,\bigl(Y^{[0]}=1\land Y^{[i+1]}=Y^{[i]}\cdot X\bigr)\]
(here it makes no difference whether we use a linear or triangular array of witnesses) and its aggregate function
version $\M{POW}^*$. Over~$\vtc$, we clearly have $\imul\vdash\M{POW}^*\vdash\M{POW}$. The
argument in \th\ref{lem:vtcim-cn}~\ref{item:10} only needed the sequence of powers $(2^n-X)^i$, $i\le m$ apart from
$\vtc$, hence it actually shows $\M{POW}\vdash\Div$. Since $\vtc+\M{POW}^*$ is a CN theory, this implies
$\vtc+\M{POW}^*=\vtcim$. In fact, one can also show that $\vtc+\M{POW}=\vtc+\Div$ by formalizing the reduction of
powering to division from~\cite{bch}. The key point is that the result of a single division is enough to reconstruct
the whole sequence of powers $X^0,\dots,X^n$, hence we do not need any aggregate functions. If $X<2^k$ and
$m=k(n+1)+1$, let $2^{nm}=(2^m-X)Q+R$ with $R<2^m-X$ using~$\Div$, write $Q=\sum_{i<n}Y^{[i]}2^{(n-1-i)m}$ with
$Y^{[i]}<2^m$, and put $Y^{[n]}=R$. Then one can show $Y^{[0]}=1$ and
\[Y^{[j]}\le2^{kj}\land\forall i<j\,Y^{[i+1]}=XY^{[i]}\]
by induction on $j\le n$. We leave the details to the interested reader.

Let us also mention that while it is unclear whether the soundness of the Hesse--Allender--Barrington
algorithms for division and iterated multiplication is provable in~$\vtc$, it seems very likely that it is provable in
$\vtc+\imul$. If true, this would imply that $\vtc+\imul$ is $\Pi^1_1$-axiomatizable over~$\vtc$ by the sentence
asserting the soundness of the algorithm, and it can be formulated as a purely universal theory in the language of
$\ob{\vtc}$. A priori, the $\imul$ axiom is only $\forall\Sig1$.

Even though we do not know whether $\imul$ is provable in~$\vtc$ itself, we can place it reasonably low in the usual
hierarchy of theories for small complexity classes: it is straightforward to show that $\vtcim$ is included in the
theory $\M{VNC}^2$ (and even $\M{VTC}^1$, if anyone bothered to define such a theory) by formalizing the computation of
iterated products by a balanced tree of binary products.

As stated in the Introduction, the provability of $\io$ in $\vtc$ or $\vtcim$ can be phrased in terms of $\tc$ root-finding
algorithms. There are several ways of expressing this connection precisely; one version reads as follows.

\begin{Prop}\th\label{lem:iop-rootf}
$\vtcim$ proves $\io$ if and only if for every constant~$d>0$ there exist $L_{\ob\vtcim}$-terms
$R_-(A_0,\dots,A_d,X,Y,E)$ and $R_+(A_0,\dots,A_d,X,Y,E)$
such that the theory proves
\begin{multline}\label{eq:9}
X<Y\land F(X)<0<F(Y)\land E>0\land Z_\pm=R_\pm(A_0,\dots,A_d,X,Y,E)\\
\to X<Z_-<Z_+<Y\land Z_+-Z_-<E\land F(Z_-)<0<F(Z_+),
\end{multline}
where all second-sort variables are interpreted as binary
rational numbers (fractions), and $F(X)$ denotes $A_dX^d+A_{d-1}X^{d-1}+\dots+A_0$.
\end{Prop}
\begin{Pf}
Left-to-right: the statement that for every $A_0,\dots,A_d,X,Y,E$ there exist $Z_-,Z_+$ satisfying~\eqref{eq:9} is provable
in~$\io$ (in the real closure of the model, there is a root of $F$ between $X$ and~$Y$ where $F$ changes sign, and this
root can be arbitrarily closely approximated from either side in the fraction field of the model using \th\ref{thm:shep}).
By assumption, the same statement is also provable in $\ob\vtcim$. Since the latter is a universal theory whose terms
are closed under definitions by cases, Herbrand's theorem implies that there are terms $R_-,R_+$ witnessing $Z_-,Z_+$.

Right-to-left: Let $D$ be a DOR induced by a model of $\vtcim$, $K$ its fraction field, and $F$ a polynomial with
coefficients in~$D$. Since $F$ can change sign only $\deg(F)$ times, a repeated use of~\eqref{eq:9} gives us elements
$Z_0<Z_1<\dots<Z_k$ of~$K$ such that $F$ has (in~$K$) a constant sign on each interval $(-\infty,Z_0)$, $(Z_k,\infty)$,
and $(Z_i,Z_{i+1})$, except when $Z_{i+1}-Z_i<1$. We have $D\model\Div$, hence we can approximate each~$Z_i$ in~$D$
within distance~$1$; it follows that in~$D$, $F$ is positive on a finite union of (possibly degenerate) intervals.
Every $L_{\M{OR}}$ open formula $\fii$ is equivalent to a Boolean combination of formulas of the form $F(X)>0$, hence
$\{X\in D:X\ge0\land\neg\fii(X)\}$ is also a finite union of intervals, and as such it has a least element if nonempty.
Thus, $D$ satisfies induction for~$\fii$.
\end{Pf}

Note that $L_{\ob\vtcim}$-terms denote $\tc$ algorithms (employing iterated multiplication), hence the gist of
the conclusion of \th\ref{lem:iop-rootf} is that $\vtcim$ proves the soundness of a $\tc$ degree-$d$ polynomial
root-approximation algorithm for each~$d$. The details can be varied; for example, we could drop $X$ and~$Y$, and make
the algorithm output approximations to all real roots of the polynomial, or even complex roots. However, such
modifications make it more difficult to state what exactly the ``soundness'' of the algorithm means.

\section{Working in $\vtcim$}\label{sec:arithmetic-vtcim}

As we already warned the reader, the objects we work with most often in this paper are binary numbers (integer or
rational), and we will employ common mathematical notation rather than the formal conventions used in~\cite{cook-ngu}:
in particular, we will typically denote numbers by lowercase letters (conversely, we will occasionally denote
unary numbers by capital letters), and we will write $x_i$ for the $i$th member of a
sequence~$x$ (which may be a constant-length tuple, a variable-length finite sequence encoded by a set as in
Section~\ref{sec:preliminaries}, or an infinite sequence given by a $\tc$~function with unary input~$i$). We do not
distinguish binary and unary numbers in notation; we will either explicitly mention which numbers are unary, or it will
be assumed from the context: unary natural numbers appear as indices and lengths of sequences, as powering exponents,
and as bound variables in iterated sums $\sum_{i=0}^nx_i$ and products $\prod_{i=0}^nx_i$.

By \th\ref{thm:cn-th}, we can use $L_{\ob\vtcim}$-function symbols (i.e., $\tc$ algorithms) freely in the arguments. In
particular, we can use basic arithmetic operations on integers, including iterated sums and products.
Iterated sums satisfy the recursive identities
\begin{align*}
\sum_{i<0}x_0&=0,\\
\sum_{i<n+1}x_i&=\sum_{i<n}x_i+x_n,
\end{align*}
and other basic properties can be easily proved by induction, for example
\begin{equation}\label{eq:10}
\begin{split}
\sum_{i<n}(x_i+y_i)&=\sum_{i<n}x_i+\sum_{i<n}y_i,\\
\sum_{i<n}yx_i&=y\sum_{i<n}x_i,\\
\sum_{i<n+m}x_i&=\sum_{i<n}x_i+\sum_{i<m}x_{n+i}.
\end{split}
\end{equation}
In particular, $\vtcim$ proves that if $\pi$ is a permutation of
$\{0,\dots,n-1\}$, then
\begin{equation}\label{eq:11}
\sum_{i<n}x_i=\sum_{i<n}x_{\pi(i)}.
\end{equation}
(In order to see this, show $\sum_{i<m}x_i=\sum_{i<n}x_{\pi(i)}[\pi(i)<m]$ by induction on~$m\le n$
using~\eqref{eq:10}, where $[\cdots]$ denotes the Iverson bracket.) This allows us to make sense of more general sums
$\sum_{i\in I}x_i$ where the indices run over a $\tc$-definable collection of objects (e.g., tuples of unary numbers)
that can be enumerated by a subset of some $\{0,\dots,n-1\}$; the identity~\eqref{eq:11} shows that the value of such a
sum is independent of the enumeration. For example, we can write
\[f(n)=\sum_{i+j=n}x_{i,j},\]
meaning a sum over all pairs of numbers $\p{i,j}$ such that $i+j=n$. We can also prove the double counting identity
\begin{equation}\label{eq:12}
\sum_{i<n}\sum_{j<m}x_{i,j}=\sum_{\substack{i<n\\j<m}}x_{i,j}=\sum_{j<m}\sum_{i<n}x_{i,j}
\end{equation}
by first showing $\sum_{i<n}\sum_{j<m}x_{i,j}=\sum_{k<nm}x_{\fl{k/m},k\bmod m}$ by induction on~$n$
using~\eqref{eq:10}, and then \eqref{eq:11} implies that other enumerations of the same set of pairs give the same
result. Likewise, we can show
\begin{equation}\label{eq:13}
\Bigl(\sum_{i<n}x_i\Bigr)\Bigl(\sum_{i<m}y_i\Bigr)=\sum_{\substack{i<n\\j<m}}x_iy_j.
\end{equation}

Iterated products can be treated the same way as sums, mutatis mutandis.

Rational numbers can be represented in $\vtcim$ as pairs of integers standing for fractions $a/b$, where~$b>0$. We will
not assume fractions to be reduced, as we cannot compute integer gcd. Arithmetic operations can be extended to rational
numbers in $\vtcim$ in the obvious way, for example
\[\sum_{i<n}\frac{a_i}{b_i}:=\frac{\sum_{i<n}a_i\prod_{j\ne i}b_j}{\prod_{i<n}b_i}.\]
$\vtcim$ knows the rationals form an ordered field, being the fraction field of a DOR. The properties of iterated sums
and products we established above for integers also hold for rationals.

Using iterated products, we can define factorials and binomial coefficients
\[n!=\prod_{i=1}^ni,\qquad\binom nm=\frac{n!}{m!(n-m)!}\]
for unary natural numbers $n\ge m$. A priori, $n!$ is a binary integer, and $\binom nm$ a binary rational; however, the
definition easily implies the identities
\[\binom n0=\binom nn=0,\qquad\binom{n+1}{m+1}=\binom nm+\binom n{m+1},\]
from which one can show by induction on~$n$ that $\binom nm$ is an integer for all $m\le n$. We can also prove by
induction on~$n$ the binomial formula
\[(x+y)^n=\sum_{i\le n}\binom nix^iy^{n-i}\]
for rational $x,y$. More generally, we can define the multinomial coefficients
\[\binom n{n_1,\dots,n_d}=\frac{n!}{n_1!\cdots n_d!}=\binom n{n_1}\binom{n-n_1}{n_2}\cdots\binom{n-n_1-\dots-n_{d-1}}{n_d}\]
for a standard constant~$d$ and unary $n=n_1+\dots+n_d$, and we can prove the multinomial formula
\begin{equation}\label{eq:14}
(x_1+\dots+x_d)^n=\sum_{n_1+\dots+n_d=n}\binom n{n_1,\dots,n_d}x_1^{n_1}\cdots x_d^{n_d}
\end{equation}
by metainduction on~$d$.

\section{Lagrange inversion formula}\label{sec:lagr-invers-form}

The Lagrange inversion formula (LIF) is an expression for the coefficients of the (compositional) inverse $g=f^{-1}$ of
a power series~$f$. In this section, we will formalize in $\vtcim$ variants of LIF for the special case where $f$ is a
constant-degree polynomial; we first show that $g$ inverts~$f$ as a formal power series, and then with the help of a
suitable bound on the coefficients of~$g$, we show that the series~$g(w)$ is convergent for small enough~$w$; this
means that under some restrictions, partial sums of $g(-a_0)$ approximate a root of the polynomial $f(x)+a_0$.

LIF, specifically the equivalent identity~\eqref{eq:cm}, has a simple combinatorial interpretation in terms of trees which allows for a straightforward bijective proof. However, this proof relies on exact counting of exponentially
many objects, and as such it cannot be formalized in $\vtcim$. In contrast, the inductive proof we give below proceeds by
low-level manipulations of sums and products; while it lacks conceptual clarity, it is elementary enough to go through
in our weak theory.

We introduce some notation for convenience. Let us fix a standard constant $d\ge1$. We are going to work
extensively with sequences $m=\p{m_2,\dots,m_d}$ of length~$d-1$ of unary nonnegative integers. We will use subscripts
$i=2,\dots,d$ to extract elements of the sequence as indicated, and we will employ superscripts (and primes) to
label various sequences used at the same time; these do not denote exponentiation.
If $m^1$ and~$m^2$ are
two such sequences, we define $m^1+m^2$ and $m^1-m^2$ coordinatewise (i.e., $(m^1+m^2)_i=m^1_i+m^2_i$), we write
$m^1\le m^2$ if $m^1_i\le m^2_i$ for all $i=2,\dots,d$, and $m^1\lneq m^2$ if $m^1\le m^2$ and $m^1\ne m^2$. We define
the generalized Catalan numbers
\[C_m=\frac{\bigl(\sum_{i=2}^dim_i\bigr)!}
        {\bigl(\sum_{i=2}^d(i-1)m_i+1\bigr)!\prod_{i=2}^dm_i!}.\]

\begin{Thm}\th\label{thm:lif}
$\vtcim$ proves the following for every constant~$d\ge1$:
let
\[f(x)=x+\sum_{k=2}^da_kx^k\]
be a rational polynomial, and let
\[g(w)=\sum_{n=1}^\infty b_nw^n\]
be the formal power series (with unary indices) defined by
\begin{equation}\label{eq:15}
b_n=\sum_{\sum_i(i-1)m_i=n-1}C_m\prod_{i=2}^d(-a_i)^{m_i}.
\end{equation}
Then $f(g(w))=w$ as formal power series.
\end{Thm}
\begin{Rem}
The sum in~\eqref{eq:15} runs over sequences $m=\p{m_2,\dots,m_d}$ satisfying the constraint $\sum_{i=2}^d(i-1)m_i=n-1$;
since this implies $m_2,\dots,m_d<n$, there are at most $n^{d-1}$ such sequences, hence the sum makes sense in $\vtcim$.

The power series identity $f(g(w))=w$ in the conclusion of the theorem amounts to $b_1=1$, and the recurrence
\begin{equation}\label{eq:ab}
b_n=\sum_{k=2}^d(-a_k)\sum_{n_1+\dots+n_k=n}b_{n_1}\cdots b_{n_k}
\qquad(n>1).
\end{equation}
Rather than developing a general theory of formal power series in $\vtcim$, we take this as a \emph{definition} of
$f(g(w))=w$.
\end{Rem}
\begin{Pf}
After plugging in the definition of~$b_n$, both sides of~\eqref{eq:ab} can be written as polynomials in
$-a_2,\dots,-a_d$ with rational (actually, integer) coefficients by several applications of~\eqref{eq:13}. Moreover,
$b_{n_j}$ contains only monomials~$\prod_i(-a_i)^{m^j_i}$ with
$\sum_i(i-1)m^j_i=n_j-1$. Thus, the right-hand side contains monomials
$\prod_i(-a_i)^{m_i}$ with $m_i=\sum_jm^j_i+\delta^k_i$, where
$\delta^k_i$ is Kronecker's delta. We have
$\sum_i(i-1)m_i=\sum_{i,j}(i-1)m^j_i+k-1=\sum_j(n_j-1)+k-1=n-1$, which
is the same constraint as on the left-hand side. In order to prove
\eqref{eq:ab}, it thus suffices to show that the coefficients of the
monomials~$\prod_i(-a_i)^{m_i}$ satisfying $\sum_i(i-1)m_i=n-1$ are the same on both sides of~\eqref{eq:ab}. This is
easily seen to be equivalent to the following identity for every
sequence~$m$:
\begin{equation}\label{eq:cm}
C_m=\sum_{k=2}^d\sum_{m^1+\dots+m^k=m-\delta^k}C_{m^1}\cdots C_{m^k}
\qquad(m\ne\vec0).
\end{equation}
(Here, we treat Kronecker's delta as the sequence $\delta^k=\p{\delta^k_2,\dots,\delta^k_d}$.)
We will prove~\eqref{eq:cm} by induction on~$\sum_im_i$,
simultaneously with the identities
\begin{align}
\label{eq:bin}
\sum_{m'+m''=m}\bigl(\tsum_i(i-1)m'_i+1\bigr)C_{m'}C_{m''}&=
  \bigl(\tsum_iim_i+1\bigr)C_m,\\
\label{eq:k}
\sum_{m^1+\dots+m^k=m}C_{m^1}\cdots C_{m^k}&=
  \frac{\bigl(\sum_iim_i+k-1\bigr)!\,k}
       {\bigl(\sum_i(i-1)m_i+k\bigr)!\prod_im_i!}
  \qquad(k=1,\dots,d).
\end{align}
The reader may find it helpful to consider the following combinatorial explanation of the identities, even though it
cannot be expressed in $\vtcim$. First, $C_m$ counts the number of ordered rooted trees with $m_2,\dots,m_d$ nodes of
out-degree $2,\dots,d$, respectively, and the appropriate number (i.e., $\sum_i(i-1)m_i+1$) of leaves. Indeed, such a tree
can be uniquely described by the sequence of out-degrees of its nodes in preorder. One checks easily that every string with
$m_2,\dots,m_d$ occurrences of $2,\dots,d$, resp., and $\sum_i(i-1)m_i+1$ occurrences of~$0$, has a unique cyclic
shift that is a valid representation of a tree, so there are
\[\frac1{\sum_iim_i+1}\binom{\sum_iim_i+1}{\sum_i(i-1)m_i+1,m_2,\dots,m_d}=C_m\]
such trees. The left-hand side of~\eqref{eq:k} thus counts $k$-tuples of trees with a prescribed total number of nodes of
out-degree $2,\dots,d$; a similar argument as above shows their number equals the right-hand side (every string with
the appropriate number of symbols of each kind has exactly $k$ cyclic shifts that are concatenations of
representations of $k$~trees). The main identity~\eqref{eq:cm} expresses that a tree with more than one node can be uniquely
decomposed as a root of out-degree $k=2,\dots,d$ followed by a $k$-tuple of trees. Finally, \eqref{eq:bin} expresses that a
pair of trees $t',t''$ together with a distinguished leaf $x$ of~$t'$ uniquely represent a tree~$t$ with a
distinguished node~$x$, namely the tree obtained by identifying the root of~$t''$ with~$x$.

Let us proceed with the formal proof by induction. Assume that \eqref{eq:cm}, \eqref{eq:bin}, and~\eqref{eq:k} hold for all $m'$ such that~$m'\lneq m$, we will prove them for~$m$.

\eqref{eq:cm}: If $m\ne\vec0$, we have
\begin{align*}
\sum_{k=2}^d\sum_{m^1+\dots+m^k=m-\delta^k}C_{m^1}\cdots C_{m^k}
&=\sum_{\substack{k=2\\m_k>0}}^d
    \frac{\bigl(\sum_iim_i-1\bigr)!\,k}
         {\bigl(\sum_i(i-1)m_i+1\bigr)!\prod_{i\ne k}m_i!\,(m_k-1)!}\\
&=\frac{\bigl(\sum_iim_i-1\bigr)!}
         {\bigl(\sum_i(i-1)m_i+1\bigr)!\prod_im_i!}
  \sum_{\substack{k=2\\m_k>0}}^dkm_k\\
&=\frac{\bigl(\sum_iim_i\bigr)!}
         {\bigl(\sum_i(i-1)m_i+1\bigr)!\prod_im_i!}
=C_m,
\end{align*}
using \eqref{eq:k} for~$m-\delta^k\lneq m$.

\eqref{eq:bin}: If $m=\vec0$, the statement holds. Otherwise, we have
\begin{align*}
\bigl(\tsum_iim_i&+1\bigr)C_m\\
&=C_m+\bigl({\tsum_iim_i}\bigr)
  \sum_{k=2}^d\sum_{m^1+\dots+m^k=m-\delta^k}C_{m^1}\cdots C_{m^k}\\
&=C_m+\sum_{k=2}^d\sum_{m^1+\dots+m^k=m-\delta^k}
  \sum_{j=1}^k\bigl({\tsum_iim^j_i+1}\bigr)C_{m^1}\cdots C_{m^k}\\
&=C_m+\sum_{k=2}^dk\sum_{m^1+\dots+m^k=m-\delta^k}
  \bigl({\tsum_iim^k_i+1}\bigr)C_{m^1}\cdots C_{m^k}\numberthis\label{eq:sym}\\
&=C_m+\sum_{k=2}^dk\sum_{m^1+\dots+m^k+m''=m-\delta^k}
  \bigl({\tsum_i(i-1)m^k_i+1}\bigr)C_{m^1}\cdots C_{m^k}C_{m''}\\
&=C_m+\sum_{\substack{m'+m''=m\\m'\ne\vec0}}C_{m''}\sum_{k=2}^dk\sum_{m^1+\dots+m^k=m'-\delta^k}
  \bigl({\tsum_i(i-1)m^k_i+1}\bigr)C_{m^1}\cdots C_{m^k}\\
&=C_m+\sum_{\substack{m'+m''=m\\m'\ne\vec0}}C_{m''}\sum_{k=2}^d\sum_{m^1+\dots+m^k=m'-\delta^k}
  \sum_{j=1}^k\bigl({\tsum_i(i-1)m^j_i+1}\bigr)C_{m^1}\cdots C_{m^k}\numberthis\label{eq:symrev}\\
&=C_m+\sum_{\substack{m'+m''=m\\m'\ne\vec0}}C_{m''}\sum_{k=2}^d\sum_{m^1+\dots+m^k=m'-\delta^k}
  \bigl({\tsum_i(i-1)m'_i+1}\bigr)C_{m^1}\cdots C_{m^k}\\
&=C_m+\sum_{\substack{m'+m''=m\\m'\ne\vec0}}\bigl({\tsum_i(i-1)m'_i+1}\bigr)C_{m''}C_{m'}\\
&=\sum_{m'+m''=m}\bigl({\tsum_i(i-1)m'_i+1}\bigr)C_{m'}C_{m''},
\end{align*}
using \eqref{eq:cm} for $m$ and~$m'\le m$, and \eqref{eq:bin} for~$m^k\lneq m$. We derive line~\eqref{eq:sym}
by observing that the $k$~sums
\[\sum_{m^1+\dots+m^k=m-\delta^k}\bigl({\tsum_iim^j_i+1}\bigr)C_{m^1}\cdots C_{m^k}\qquad(j=1,\dots,k)\]
have the same value due to symmetry (i.e., by an application of~\eqref{eq:11}).
Line~\eqref{eq:symrev} is similar.

\eqref{eq:k}: By metainduction on~$k=1,\dots,d$. The case~$k=1$
is the definition of~$C_m$. Assuming the statement holds
for~$k$, we prove it for~$k+1$ from the identity
\begin{align*}
k\bigl(\tsum_i(i&-1)m_i+k+1\bigr)
  \sum_{m^1+\dots+m^{k+1}=m}C_{m^1}\cdots C_{m^{k+1}}\\
&=k\sum_{m^1+\dots+m^{k+1}=m}\sum_{j=1}^{k+1}
    \bigl(\tsum_i(i-1)m^j_i+1\bigr)C_{m^1}\cdots C_{m^{k+1}}\\
&=k(k+1)\sum_{m^1+\dots+m^{k+1}=m}
    \bigl(\tsum_i(i-1)m^{k+1}_i+1\bigr)C_{m^1}\cdots C_{m^{k+1}}\\
&=k(k+1)\sum_{m^1+\dots+m^k=m}C_{m^1}\cdots C_{m^{k-1}}
    \sum_{m'+m''=m^k}\bigl(\tsum_i(i-1)m'_i+1\bigr)C_{m'}C_{m''}\\
&=k(k+1)\sum_{m^1+\dots+m^k=m}
    \bigl(\tsum_iim^k_i+1\bigr)C_{m^1}\cdots C_{m^k}\\
&=(k+1)\sum_{m^1+\dots+m^k=m}\sum_{j=1}^k
    \bigl(\tsum_iim^j_i+1\bigr)C_{m^1}\cdots C_{m^k}\\
&=(k+1)\bigl({\tsum_iim_i+k}\bigr)\sum_{m^1+\dots+m^k=m}
    C_{m^1}\cdots C_{m^k}
\end{align*}
using \eqref{eq:bin} for~$m^k\le m$.
\end{Pf}

\begin{Lem}\th\label{lem:coefbd}
$\vtcim$ proves:
let $f,g$ be as in \th\ref{thm:lif}, and
$a=\max\bigl\{1,\sum_i\abs{a_i}\bigr\}$. Then $\abs{b_n}\le(4a)^{n-1}$ for
every~$n$.
\end{Lem}
\begin{Pf}
We can estimate
\begin{align*}
\abs{b_n}
&\le a^{n-1}\sum_{\sum_i(i-1)m_i=n-1}C_m
                    \prod_{i=2}^d\bigl(a^{1-i}\abs{a_i}\bigr)^{m_i}\\
&=\frac{a^{n-1}}n\sum_{\sum_i(i-1)m_i=n-1}
          \binom{n-1+\sum_im_i}{n-1,m_2,\dots,m_d}
          \prod_{i=2}^d\bigl(a^{1-i}\abs{a_i}\bigr)^{m_i}\\
&\le\frac{a^{n-1}}n\sum_{t=n-1}^{2(n-1)}
        \sum_{s+\sum_im_i=t}\binom t{s,m_2,\dots,m_d}
          \prod_{i=2}^d\bigl(a^{-1}\abs{a_i}\bigr)^{m_i}\\
&=\frac{a^{n-1}}n\sum_{t=n-1}^{2(n-1)}\Bigl(1+a^{-1}\sum_{i=2}^d\abs{a_i}\Bigr)^t\\
&\le a^{n-1}\Bigl(1+a^{-1}\sum_{i=2}^d\abs{a_i}\Bigr)^{2(n-1)}
\le a^{n-1}2^{2(n-1)}
\end{align*}
using the multinomial formula~\eqref{eq:14}.
\end{Pf}

\begin{Exm}
The bound in \th\ref{lem:coefbd} is reasonably tight even in the ``real world''.
Let $a>0$ be a real number, and put $f(x)=x-ax^2$. Then $g$ is its
inverse function $g(w)=(1-\sqrt{1-4aw})/2a$, whose radius of
convergence is the modulus of the nearest singularity, namely
$1/4a$. Thus, for every~$\ep>0$, $\abs{b_n}\ge(4a-\ep)^n$ for
infinitely many~$n$. In fact, the Stirling approximation for Catalan numbers gives
$b_n=\Theta\bigl((4a)^nn^{-3/2}\bigr)$.
\end{Exm}

\begin{Thm}\th\label{thm:lifnum}
$\vtcim$ proves the following for every constant~$d\ge1$.
Let $h(x)=\sum_{i=0}^da_ix^i$ be a rational polynomial with linear
coefficient $a_1=1$. Put $f=h-a_0$,
let $g$ and~$b_n$ be as in \th\ref{thm:lif},
$a=\max\bigl\{1,\sum_{i=2}^d\abs{a_i}\bigr\}$, $\alpha=4a\abs{a_0}$, and let
\[x_N=\sum_{n=1}^Nb_n(-a_0)^n\]
denote the $N$th partial sum of~$g(-a_0)$
for every unary natural number~$N$. If
\[\abs{a_0}<\frac1{4a},\]
then
\begin{align}
\label{eq:2}\abs{x_N}&\le\frac{\abs{a_0}}{1-\alpha},\\
\label{eq:5}\abs{x_N-x_M}&\le\frac{\abs{a_0}\alpha^{N-1}}{1-\alpha},\\
\label{eq:3}\abs{h(x_N)}&\le N^d\abs{a_0}\alpha^N
\end{align}
for every unary~$M\ge N\ge1$.
\end{Thm}
\begin{Pf}
\th\ref{lem:coefbd} gives
\[\abs{x_N}\le\sum_{n=1}^N\abs{a_0}^n(4a)^{n-1}
=\abs{a_0}\sum_{n=0}^{N-1}\alpha^n
\le\frac{\abs{a_0}}{1-\alpha}.\]
The proof of~\eqref{eq:5} is similar. As for~\eqref{eq:3}, we have
\begin{align}
h(x_N)
&=a_0+\sum_{k=1}^da_k\sum_{n_1,\dots,n_k=1}^N
        b_{n_1}\cdots b_{n_k}(-a_0)^{n_1+\cdots+n_k}\nonumber\\
&=\sum_{k=1}^da_k\sum_{\substack{n_1,\dots,n_k=1\\n_1+\dots+n_k>N}}^N
        b_{n_1}\cdots b_{n_k}(-a_0)^{n_1+\cdots+n_k},\label{eq:4}
\end{align}
as
\[\sum_{k=1}^da_k\sum_{n_1+\dots+n_k=n}b_{n_1}\cdots b_{n_k}=\delta^1_n\]
for all $n\le N$ by \th\ref{thm:lif}. Note that the inner sum in~\eqref{eq:4} is
empty for~$k=1$, thus
\begin{align*}
\abs{h(x_N)}&\le
\sum_{k=2}^d\abs{a_k}
   \sum_{\substack{n_1,\dots,n_k=1\\n_1+\dots+n_k>N}}^N
       (4a)^{-k}\bigl(4a\abs{a_0}\bigr)^{n_1+\cdots+n_k}\\
&\le\sum_{k=2}^d\abs{a_k}\left(\frac N{4a}\right)^k
          \alpha^{N+1}\\
&\le a\max\left\{\frac{N^2}{(4a)^2},\frac{N^d}{(4a)^d}\right\}\alpha^{N+1}\\
&\le\max\left\{\frac{N^2}{4},\frac{N^d}{4^{d-1}}\right\}\abs{a_0}\alpha^N
\le N^d\abs{a_0}\alpha^N,
\end{align*}
using \th\ref{lem:coefbd} and~$a\ge1$.
\end{Pf}
Intuitively, the conclusion of \th\ref{thm:lifnum} says that $x_N$ is
a Cauchy sequence with an explicit modulus of convergence whose
limit is a root of~$h$ of bounded modulus.
%

\section{Valued fields}\label{sec:valued-fields}

\th\ref{thm:lifnum} shows that $\vtcim$ can compute roots of
polynomials of a special form, however it would still be rather difficult to
extend it to a full-blown root-finding algorithm. We will instead give
a model-theoretic argument using well-known properties of valued
fields to bridge the gap between \th\ref{thm:lifnum} and approximation
of roots of general polynomials.

In order to prove $\vtcim\vdash\io$, it suffices to show that every
model of~$\vtcim$ is a model of~$\io$. First, since
$\vtcim\vdash\Div$, we can reformulate \th\ref{thm:shep}
in terms of fields.
\pagebreak[2]
\begin{Lem}\th\label{lem:shep-field}
Let $D$ be a DOR, and $F$ its fraction field. The following are
equivalent.
\begin{enumerate}
\item $D\model\io$.
\item $D\model\Div$, and $F$ is a dense subfield of a
RCF~$R$.\noproof
\end{enumerate}
\end{Lem}
The condition that $F$ is dense in~$R$ means that elements of~$R$
can be well approximated in~$F$, i.e., $R$ cannot be too large, while the
condition that $R$ is real-closed (or at least contains the
real closure $\rcl F$) means that $R$ cannot be too small, so these
two conditions work against each other. One canonical choice of~$R$ is
the smallest RCF extending~$F$, i.e., $\rcl F$. We obtain that a DOR $D\model\Div$ is a
model of~$\io$ iff $F$ is dense in~$\rcl F$. However, it will be
useful for us to consider another choice: it turns
out that there exists the largest ordered
field extension~$\hat F\Sset F$ in which~$F$ is dense, and
a DOR $D\model\Div$ is a model of~$\io$ iff
$\hat F$ is real-closed.

The existence of~$\hat F$ was shown by Scott~\cite{scott-cof}.
One way to prove it is by generalization of the
construction of~$\RR$ using Dedekind cuts. Consider pairs $\p{A,B}$,
where $F=A\cup B$, $B$ has no smallest element, and
\[\inf\{b-a:a\in A,b\in B\}=0.\]
One can show that the collection of all such cuts can be given the
structure of an ordered field in a natural way, and it has the
property needed of~$\hat F$. However, we will use a different
construction of~$\hat F$ which may look more complicated on first
sight, but has the advantage of allowing us to employ tools from
the theory of valuations to explore its properties (such as being
real-closed). It can be thought of as generalizing the construction
of~$\RR$ by means of Cauchy sequences.

We refer the reader to~\cite{engl-pres} for the theory of valued
fields, however we will review our notation and some basic facts
below to make sure we are on the same page.

A \emph{valuation} on a field~$K$ is a surjective mapping $v\colon
K\onto\Gamma\cup\{\infty\}$, where $\p{\Gamma,+,\le}$ is a totally
ordered abelian group (called the \emph{value group}), and $v$ satisfies
\begin{enumerate}
\item $v(a)=\infty$ only if $a=0$,
\item $v(ab)=v(a)+v(b)$,
\item $v(a+b)=\min\{v(a),v(b)\}$,
\end{enumerate}
where we put $\infty+\gamma=\gamma+\infty=\infty$ and $\gamma\le\infty$
for every~$\gamma\in\Gamma$. (Elements with large valuation should be
thought of as being small; the order is upside down for historical
reasons.) Valuations $v\colon K\to\Gamma\cup\{\infty\}$, $v'\colon
K\to\Gamma'\cup\{\infty\}$ are \emph{equivalent} if there is an
ordered group isomorphism $f\colon\Gamma\to\Gamma'$ such that
$v'=f\circ v$.

The \emph{valuation ring} of $v$ is
\[O=\{a\in K:v(a)\ge0\},\]
with its unique \emph{maximal ideal} being
\[I=\{a\in K:v(a)>0\}.\]
The quotient field~$k=O/I$ is called the \emph{residue field}. If
$a\in O$, we will denote its image under the natural projection~$O\to
k$ as~$\ob a$.

More abstractly, a valuation ring for a field~$K$ is a subring $O\sset
K$ such that $a\in O$ or~$a^{-1}\in O$ for every~$a\in K^\times$. Any such
ring corresponds to a valuation: we take $\Gamma=K^\times/O^\times$ ordered by
$aO^\times\le bO^\times$ iff $b\in aO$, and define~$v$ as the natural projection
$v(a)=aO^\times$. A valuation is determined uniquely up to equivalence by
its valuation ring; thus, either of the structures $\p{K,v}$
and~$\p{K,O}$ can be called a \emph{valued field}. A valued
field~$\p{K',v'}$ is an \emph{extension} of~$\p{K,v}$ if $K$ is a
subfield of~$K'$, and $v\sset v'$. (In terms of valuation rings, the
latter means $O=O'\cap K$.) A valuation (or valuation ring or valued
field) is \emph{nontrivial} if~$\Gamma\ne\{0\}$, or equivalently, if~$O\ne K$.

A valuation $v\colon K\to\Gamma\cup\{\infty\}$ induces a
\emph{topology} on~$K$ with basic open sets
\[B(a,\gamma)=\{b\in K:v(b-a)>\gamma\},\qquad a\in K, \gamma\in\Gamma.\]
(Note that $B(a,\gamma)=B(a',\gamma)$ for any~$a'\in B(a,\gamma)$.)
This makes $K$ a topological field, and as with any topological group,
it also makes $K$ a uniform space (with a fundamental system of entourages
of the form $\{\p{a,b}\in K^2:v(a-b)>\gamma\}$
for~$\gamma\in\Gamma$). Consequently, we have the notions of Cauchy nets,
completeness, and completion; for the particular case of valued
fields, they can be stated as follows. A \emph{Cauchy sequence} in~$K$
is
$\{a_\gamma:\gamma\in\Gamma\}\sset K$ such that
$v(a_\gamma-a_\delta)>\min\{\gamma,\delta\}$ for
every~$\gamma,\delta\in\Gamma$. (Alternatively, it would be enough if
Cauchy sequences were indexed over a cofinal subset of~$\Gamma$.) Such
a sequence converges to~$a\in K$ if $v(a-a_\gamma)>\gamma$ for
every~$\gamma\in\Gamma$. The valued field~$\p{K,v}$ is \emph{complete}
if every Cauchy sequence in~$K$ converges. A \emph{completion}
of~$\p{K,v}$ is an extension~$\p{\hat K,\hat v}$ of~$\p{K,v}$ which is
a complete valued field such that $K$ is (topologically) dense
in~$\hat K$. (The last condition implies that
$\hat K$ is an \emph{immediate} extension of~$K$, i.e.,
the natural embeddings $\Gamma\sset\hat\Gamma$ and $k\sset\hat k$ are
isomorphisms.)
\begin{Thm}[{{\cite[Thm.2.4.3]{engl-pres}}}]\th\label{thm:compl}
Every valued field~$\p{K,v}$ has a completion, which is unique up to a
unique valued field isomorphism identical on~$K$.
\noproof\end{Thm}

Now we turn to the interaction of valuation and order~\cite[\S2.2.2]{engl-pres}. Let
$\p{K,O}$ be a valued field. If $\le$ is an order on~$K$ (i.e.,
$\p{K,\le}$ is an ordered field) such that $O$ is \emph{convex} (i.e.,
$a\le b\le c$ and~$a,c\in O$ implies~$b\in O$), then an order is
induced on the residue field~$k$ by $\ob a\le\ob b\EQ a\le b$.
Conversely, any order on~$k$ is induced from an order~$\le$ on~$K$
making $O$ convex in this way. If $\Gamma$ is $2$-divisible, such a~$\le$ is
unique, and can be defined explicitly by
\[a>0\iff\exists b\in K\,(ab^2\in O^\times\land\ob{ab^2}>0).\]
In general, the structure of all such orders~$\le$ is described by the Baer--Krull
theorem \cite[Thm.~2.2.5]{engl-pres}. Notice also that every convex
subring of an ordered field is a valuation ring.
\begin{Lem}\th\label{lem:toptop}
If $\p{K,\le}$ is an ordered field, and $O$ a nontrivial convex subring of~$K$, then the valuation topology on~$K$
coincides with the interval topology. In particular, a
subset~$X\sset K$ is topologically dense iff it is order-theoretically
dense.
\end{Lem}
\begin{Pf}
The convexity of~$O$ implies that every~$B(a,\gamma)$ is also convex.
If $c\in(a,b)$, and $\gamma\ge v(c-a),v(c-b)$, then $c\in
B(c,\gamma)\sset(a,b)$. On the other hand, if $c\in B(a,\gamma)$, pick
$e>0$ with~$v(e)>\gamma$ (which exists as the valuation is
nontrivial). Then $c\in(c-e,c+e)\sset B(a,\gamma)$.
\end{Pf}
For any ordered field $\p{K,\le}$, the set of its bounded elements
\[O=\{a\in K:\exists q\in\Q^+\,(-q\le a\le q)\}\]
is a convex valuation ring for~$K$ with the set of infinitesimal elements
\[I=\{a\in K:\forall q\in\Q^+\,(-q\le a\le q)\}\]
being its maximal ideal. The corresponding valuation is the
\emph{natural valuation} induced by~$\le$. The residue field is an
archimedean ordered field, and as such it can be uniquely identified
with a subfield~$k\sset\RR$. Here is the promised construction of the
largest dense extension of an ordered field.
\begin{Lem}\th\label{lem:ordcomp}
Let $\p{K,\le}$ be a nonarchimedean ordered field, $v$ its natural
valuation, and $\p{\hat K,\hat v}$ its completion. There is a
unique order on~$\hat K$ extending~$\le$ that makes~$\hat O$ convex.
Its natural valuation is~$\hat v$, and it satisfies:
\begin{enumerate}
\item\label{item:1} $\hat K$ is an ordered field extension
of~$K$ such that $K$ is dense in~$\hat K$.
\item\label{item:2} If $K'$ is any ordered field extension of~$K$ in which
$K$ is dense, there is a unique ordered field embedding of~$K'$
in~$\hat K$ identical on~$K$.
\end{enumerate}
\end{Lem}
\begin{Pf}
Since $\hat K$ is an immediate extension of~$K$, for every~$a\in\hat
K^\times$ there exists an~$a_0\in K^\times$ such that $aa_0^{-1}\in1+\hat I$, or
equivalently, $\hat v(a-a_0)>\hat v(a)=v(a_0)$. Any order~$\hat\le$
on~$\hat K$ extending~$\le$ such that $\hat O$ is convex (which
implies $1+\hat I\sset\hat K^+$) must satisfy
\begin{equation}\label{eq:6}
a\mr{\hat>}0\iff a_0>0,
\end{equation}
which specifies it uniquely. On the other hand, we claim
that~\eqref{eq:6} defines an order on~$\hat K$. First, the definition
is independent of the choice of~$a_0$: if $a_1\in K^\times$ is such that
$aa_1^{-1}\in1+\hat I$, then $a_0a_1^{-1}\in1+I$ is positive, whence
$a_0$ and~$a_1$ have the same sign. Clearly, exactly one of $a$
and~$-a$ is positive for any~$a\in \hat K^\times$. Let $a,b\in\hat K^\times$,
$a,b\mr{\hat>}0$. Since $(ab)(a_0b_0)^{-1}\in1+\hat I$, we have
$ab\mr{\hat>}0$. Also, $v(a_0+b_0)=\min\{v(a_0),v(b_0)\}$ as they have
the same sign, thus
\[\hat v\bigl((a+b)-(a_0+b_0)\bigr)
\ge\min\{\hat v(a-a_0),\hat v(b-b_0)\}
>\min\{v(a_0),v(b_0)\}=v(a_0+b_0).\]
This means we can take $a_0+b_0$ for~$(a+b)_0$, showing
that~$a+b\mr{\hat>}0$.

If $a\mr{\hat<}b\mr{\hat<}c$, $a,c\in\hat O$, we may assume
$(c-a)_0=(c-b)_0+(b-a)_0$ by the argument above, hence
$(c-b)_0+(b-a)_0\in O$. Since $(c-b)_0,(b-a)_0>0$, this implies
$(b-a)_0\in O$, hence~$b-a\in\hat O$, and~$b\in\hat O$. Thus, $\hat O$
is convex under~$\hat\le$.

Since $\p{K,\le}$ is nonarchimedean, the valuations $v$ and~$\hat v$
are nontrivial. Thus, $K$ is an order-theoretically dense subfield
of~$\hat K$ by \th\ref{lem:toptop}, which shows~\ref{item:1}. Also, in view of the convexity
of~$\hat O$, this implies that $O$ is dense in~$\hat O$, hence
\[\hat O=\{a\in \hat K:\exists
q\in\Q^+\,(-q\mr{\hat\le}a\mr{\hat\le}q)\},\]
i.e., $\hat v$ is the natural valuation of~$\p{\hat K,\hat\le}$.

\ref{item:2}: Let $v'$ be the natural valuation on~$K'$, and $\p{\hat
K',\hat v'}$ its completion. By \th\ref{lem:toptop}, $\p{K,v}$ is
topologically dense in its complete extension~$\p{\hat K',\hat v'}$,
hence there is an isomorphism of~$\p{\hat K',\hat v'}$ and~$\p{\hat
K,\hat v}$ identical on~$K$ by \th\ref{thm:compl}. It restricts to an
embedding $f\colon\p{K',v'}\to\p{\hat K,\hat v}$. For any $a\in K'$,
we can see from~\eqref{eq:6} that $f(a)\mr{\hat>}0$ implies $a_0>0$ for some
$a_0\in K^\times$ such that $aa_0^{-1}\in1+I'$, whence~$a\mr{>'}0$. Thus,
$f$ is order-preserving. The uniqueness of~$f$ follows from the
density of~$K$ in~$\hat K$.
\end{Pf}
(If $K$ is archimedean, its natural valuation is trivial, hence the induced
topology is discrete, and~$\hat K=K$. However, the largest ordered
field extension of~$K$ where $K$ is dense is~$\RR$.)

We will rely on the following important characterization of real-closed fields in
terms of valuations \cite[Thm.~4.3.7]{engl-pres}.
\begin{Thm}\th\label{thm:rcfval}
Let $\p{K,\le}$ be an ordered field, and $O$ a convex valuation ring
of~$K$. The following are equivalent.
\begin{enumerate}
\item $K$ is real-closed.
\item $\Gamma$ is divisible, $k$ is real-closed, and $O$ is henselian.\noproof
\end{enumerate}
\end{Thm}
There are many equivalent definitions of henselian valuation rings or
valued fields (cf.~\cite[Thm.~4.1.3]{engl-pres}). It will be most
convenient for our purposes to adopt the following one: a valuation
ring~$O$ or a valued field~$\p{K,O}$ is \emph{henselian} iff every polynomial
$h(x)=\sum_{i=0}^da_ix^i\in O[x]$ such that $a_0\in I$ and~$a_1=1$ has
a root in~$I$.

The basic intuition behind \th\ref{thm:rcfval} is that in order to
find a root~$a$ of a polynomial in~$K$, we use the divisibility
of~$\Gamma$ to get a ballpark estimate of~$a$, we refine it to an
approximation up to an infinitesimal relative error using the
real-closedness of~$k$, and then use the henselian property to
compute~$a$. Complications arise from interference with other roots of the
polynomial.

It is well known that the completion of a henselian valued field is
henselian. In fact, we have the following simple criterion, where we
define a valued field~$\p{K,O}$ to be \emph{almost henselian}
if for every polynomial~$h$ as above, and every~$\gamma\in\Gamma$,
there is~$a\in I$ such that~$v(h(a))>\gamma$. (Equivalently, $\p{K,O}$
is almost henselian iff the quotient ring~$O/P$ is henselian for every
nonzero prime ideal~$P\sset O$ \cite{vamos}.)
\begin{Lem}\th\label{lem:apxhen}
The completion~$\p{\hat K,\hat v}$ is henselian iff $\p{K,v}$ is
almost henselian.
\end{Lem}
\begin{Pf}
First, we observe that if $h=\sum_{i=0}^da_ix^i\in O[x]$ has~$a_1=1$,
then
\begin{equation}\label{eq:7}
v(h(b)-h(c))=v(b-c)
\end{equation}
for any~$b,c\in I$. Indeed, if~$b\ne c$, we have
\[\frac{h(b)-h(c)}{b-c}=a_1+\sum_{i=2}^da_i(b^{i-1}+b^{i-2}c+\dots+c^{i-1})
\in1+I\sset O^\times.\]

Left to right: assume that $h=\sum_{i=0}^da_ix^i\in O[x]$, $a_1=1$,
$a_0\in I$, and~$\gamma\in\Gamma$. Without loss of generality,
$\gamma\ge0$. Since $\hat K$ is henselian, there is~$\hat a\in\hat I$
such that~$h(\hat a)=0$. By the density of~$K$ in~$\hat K$, we can
find~$a\in K$ such that $\hat v(a-\hat a)>\gamma$. Then $a\in I$, and
$v(h(a))>\gamma$ by~\eqref{eq:7}.

Right to left: let $h=\sum_{i=0}^da_ix^i\in\hat O[x]$ with $a_1=1$
and~$a_0\in\hat I$. For any~$\gamma\in\Gamma$, $\gamma\ge0$, we choose
$a_{i,\gamma}\in K$ such that $\hat v(a_i-a_{i,\gamma})>\gamma$, and
put $h_\gamma=\sum_ia_{i,\gamma}x^i$. Then $h_\gamma\in O[x]$,
$a_{0,\gamma}\in I$, and we could have picked~$a_{1,\gamma}=1$, hence by
assumption, there is~$b_\gamma\in I$ such
that~$v(h_\gamma(b_\gamma))>\gamma$. By the choice of~$h_\gamma$, this
implies~$\hat v(h(b_\gamma))>\gamma$. Moreover,
$v(b_\gamma-b_\delta)=\hat
v(h(b_\gamma)-h(b_\delta))>\min\{\gamma,\delta\}$ by~\eqref{eq:7},
hence $\{b_\gamma:\gamma\ge0\}$ is a Cauchy sequence. Since $\hat K$
is complete, there is~$b\in\hat K$ such that $\hat
v(b-b_\gamma)>\gamma$ for every~$\gamma$. Then $b\in\hat I$. Since
$\hat v(h(b)-h(b_\gamma))>\gamma$ by~\eqref{eq:7}, we have $\hat
v(h(b))>\gamma$ for every~$\gamma\in\Gamma$, i.e., $h(b)=0$.
\end{Pf}
Putting all the things together, we obtain the following
characterization of open induction. We note that the fact that the
completion of a real-closed field is real-closed was shown by
Scott~\cite{scott-cof}.
\begin{Lem}\th\label{thm:iopcompl}
Let $D$ be a nonstandard DOR such that $D\model\Div$, $F$ its fraction
field endowed with its natural valuation, and $\hat F$ its completion. The
following are equivalent.
\begin{enumerate}
\item\label{item:4} $D\model\io$.
\item\label{item:5} $\hat F$ is real-closed.
\item\label{item:3} $F$ is almost henselian, its value group is divisible,
and its residue field is real-closed.
\end{enumerate}
\end{Lem}
\begin{Pf}
\ref{item:5} and~\ref{item:3} are equivalent by
\th\ref{thm:rcfval,lem:apxhen}, using the fact that $\hat F$ is an immediate extension of~$F$.

\ref{item:5}${}\to{}$\ref{item:4} follows from \th\ref{lem:shep-field}
as $F$ is dense in~$\hat F$. Conversely, assume that $F$ is a dense
subfield of a RCF~$R$. By
\th\ref{thm:rcfval}, $R$ is henselian, its value group is divisible,
and its residue field is a RCF. The completion~$\hat R$ is also henselian by
\th\ref{lem:apxhen}, and it has the same $\Gamma$ and~$k$ as~$R$,
hence it is a RCF by \th\ref{thm:rcfval}. However, the density of~$F$
in~$\hat R$ implies $\hat F\simeq\hat R$ by \th\ref{lem:ordcomp},
hence $\hat F$ is a RCF.
\end{Pf}
We remark that we could have used any nontrivial convex subring
in place of the natural valuation in \th\ref{lem:ordcomp} (any two
such valuations determine the same uniform structure by
\th\ref{lem:toptop}, which means that their completions are the same
qua topological fields, and one checks easily that they also carry
the same order). Likewise, \th\ref{thm:iopcompl} continues to hold
when $F$ is endowed with any nontrivial valuation with a convex
valuation ring; this may make a difference for verification of
condition~\ref{item:3}. Notice that such valuation rings correspond
to proper cuts (in the models-of-arithmetic sense) on~$D$ closed under multiplication.

We can now prove the main result of this paper.
\begin{Thm}\th\label{thm:vtciopen}
$\vtcim$ proves $\io$ on binary integers.
\end{Thm}
\begin{Pf}
Let $M\model\vtcim$, and $D$ be its ring of binary integers, we need
to show that $D\model\io$. We may assume without loss of generality
that $M$, and therefore $D$, is $\omega$-saturated. Since
$\vtcim\vdash\Div$, it suffices to check the conditions of
\th\ref{thm:iopcompl}~\ref{item:3}.

As we have mentioned above, the residue field~$k$ of any ordered field
under its natural valuation is a subfield of~$\RR$. The
$\omega$-saturation of~$D$ implies that every Dedekind cut on~$\Q$ is realized
by an element of~$F$, hence in fact~$k=\RR$, which is a real-closed
field.

Every element of the value group~$\Gamma$ is the difference of
valuations of two (positive) elements of~$D$. Let thus $a\in D^+$,
and $k\in\Z^+$. Put $n=\lh a-1$, which is a unary integer of~$M$
such that $2^n\le a<2^{n+1}$. Put $m=\fl{n/k}$ and $b=2^m$. Then
$b^k\le a<2^kb^k$, hence $kv(b)=v(a)$. This shows that $\Gamma$ is divisible.

Let $\gamma\in\Gamma$, and $h(x)=\sum_{i\le d}a_ix^i\in F[x]$ be such
that $v(a_i)\ge0$, $v(a_0)>0$, and $a_1=1$. Then
$a=\max\bigl\{1,\sum_{i=2}^d\abs{a_i}\bigr\}$ is bounded by a standard
integer, whereas $a_0$ is infinitesimal, thus $\alpha=4a\abs{a_0}$ is
also infinitesimal. Let $N$ be a nonstandard unary integer of~$M$ such
that $v(2^{-N})>\gamma$, and let $x_N$ be as in \th\ref{thm:lifnum}.
Then using a crude estimate,
\[\abs{h(x_N)}\le N^d\abs{a_0}\alpha^N\le2^N4^{-N}=2^{-N},\]
which means that $v(h(x_N))>\gamma$. Moreover,
$\abs{x_N}\le\abs{a_0}/(1-\alpha)$ is infinitesimal. Thus, $F$ is almost henselian.
\end{Pf}

As explained in Section~\ref{sec:imul-div}, \th\ref{thm:vtciopen} implies that for any
constant~$d$, $\vtcim$ can formalize a $\tc$ algorithm for
approximation of roots of degree~$d$ rational polynomials. The reader might find it disappointing that we
have shown its existence nonconstructively using the abstract nonsense
from this section, so let us give at least a rough idea how this
algorithm may actually look like; it is somewhat different from the one 
in~\cite{ej:polyroot}.

Clearly, one ingredient is \th\ref{thm:lifnum}, which gives an
explicit description of a $\tc$ algorithm for approximation of roots
of polynomials of a special form (small constant coefficient and large
linear coefficient). The remaining part is a reduction of general root
approximation to this special case, and this happens essentially in
\th\ref{thm:rcfval}. This theorem has a proof with a fairly
algorithmic flavour using Newton polygons (cf.\ \cite[\S2.6]{bpr},
where a similar argument is given in the special case of real Puiseux
series). The Newton polygon of a polynomial
$f(x)=\sum_{i=0}^da_ix^i\in K[x]$ is the lower convex hull of the set
of points $\{e_i=\p{i,v(a_i)}:i=0,\dots,d\}\sset\Q\times\Gamma$. 

The basic idea is as follows. Take an edge of the Newton polygon with
endpoints~$e_{i_0},e_{i_1}$. The slope of the edge is in~$\Gamma$ due
to its divisibility, hence we can replace $f(x)$ by a suitable
polynomial of the form~$af(bx)$ to ensure $v(a_{i_0})=v(a_{i_1})=0$.
Then $f\in O[x]$, its image~$\ob f\in k[x]$ has degree~$i_1$, and
the least exponent of its nonzero coefficient is~$i_0$. If we find a
nonzero root~$\ob a\in k^\times$ of~$\ob f$ of multiplicity~$m$ using the
real-closedness of~$k$, the Newton polygon of the
shifted polynomial~$f(x+a)$ will have an edge whose endpoints satisfy
$i'_0<i'_1\le m\le i_1-i_0$, since $m$ is the least exponent with a
nonzero coefficient in~$\ob f(x+\ob a)$. This is strictly shorter than the
original edge unless $\ob f$ is a constant multiple of~$x^{i_0}(x-\ob
a)^{i_1-i_0}$, which case has to be handled separately. If we set up
the argument properly, we can reduce~$f$ by such linear substitutions in
at most~$d$ steps into a polynomial whose Newton polygon has $e_0,e_1$
for vertices, and then we can apply the henselian property to find its
root in~$K$.

One can imagine that a proper $\tc$ algorithm working over~$\Q$
instead of a nonarchimedean field can be obtained along similar lines by
replacing ``infinitesimal'' with a suitable notion of ``small
enough'' (e.g., employing an approximation of~$-\log\Abs a$ as a
measure of magnitude in place of~$v(a)$). However, the details are
bound to be quite unsightly due to complications arising from the loss of
the ultrametric inequality of~$v$.

\section{Application to Buss's theories}\label{sec:buss-th}
\let\lh\abs

While $\vtcim$ does not stand much chance of proving induction for interesting classes of formulas with quantifiers in
the language of ordered rings, we will show in this section that we can do better in the richer language
$L_B=\p{0,1,+,\cdot,\le,\#,\lh x,\half x}$ of Buss's
one-sorted theories of bounded arithmetic---$\vtcim$ proves the $\rsuv$-translation of~$T^0_2$, and even
minimization for sharply bounded formulas ($\sig0$-\Min). The main tool is a description of $\sig0$-definable
sets discovered by Mantzivis~\cite{mantz}, whose variants were also given in \cite{bou-kol,kol:t02}: in essence, a
$\sig0$-definable subset of~$[0,2^n)$ can be written as a union of $n^{O(1)}$ intervals on each residue class
modulo~$2^c$, where $c$ is a standard constant. As we will see, this property can be formalized in~$\vtcim$ using
the provability of~$\io$ for the base case of polynomial inequalities, and as a consequence, our theory proves
minimization and induction for $\sig0$ formulas. (We stress that as in the case of~$\io$, these are minimization and
induction over binary numbers. Despite the same name, the schemata denoted as $\ind$ and~$\Min$ in the two-sorted
framework only correspond to $\lind$ and minimization over lengths in Buss's language, respectively.)
We will present the messier part of the argument as a normal form for $\sig0$ formulas over a weak base theory, in the
hope that this will make the result more reusable.

We will assume the reader is familiar with definitions of Buss's theories (see e.g.\ \cite{buss,book}), in
particular, with~$\bas$. Recall that a formula is sharply bounded if all its quantifiers are of the form $\exists
x\le\lh t$ or $\forall x\le\lh t$. We reserve $\sig0$ for the class of sharply bounded formulas of~$L_B$, whereas
sharply bounded formulas in an extended language $L_B\cup L'$ will be denoted $\sig0(L')$.  Let
$\bas^+$ denote the extension of Buss's $\bas$ by the axioms
\begin{gather}
\label{ax:1}x(yz)=(xy)z,\\
\label{ax:2}y\le x\to\exists z\,(y+z=x),\\
\label{ax:3}u\le\lh x\to\exists y\,(\lh y=u),\\
\label{ax:4}z<x\#y\to\lh z\le\lh x\lh y,\\
\label{ax:5}\lh x\le x.
\end{gather}
(The quantifiers in \eqref{ax:2}, \eqref{ax:3} could be bounded by~$x$, if desired.) On top of~$\bas$, axioms
\eqref{ax:1} and~\eqref{ax:2} imply the theory of nonnegative parts of discretely ordered rings, hence we can imagine
the universe is extended with negative numbers in the usual fashion. In particular, we can work with integer
polynomials. We introduce two extra functions by
\begin{align*}
x\dotminus y&=z\iff y+z=x\lor(x<y\land z=0),\\
2^{\min\{u,\lh x\}}&=z\iff z\#1=2z\land\bigl((u\le\lh x\land\lh z=u+1)\lor(u>\lh x\land\lh z=\lh x+1)\bigr).
\end{align*}
$\bas^+$ proves that $\dotminus$ and~$2^{\min\{u,\lh x\}}$ are well-defined total functions. Notice that $\bas^+$ is
universally axiomatizable in a language with $\dotminus$ and $2^{\min\{u,\lh x\}}$.
We will write $2^u$ for $2^{\min\{u,\lh x\}}$ when a self-evident value of~$x$ such that $u\le\lh x$ can be inferred
from the context (e.g., when $u$ is a sharply bounded quantified variable).

If $p$ is a polynomial with nonnegative
integer coefficients, one can construct easily a term~$t$ such that $\bas^+\vdash p(\lh{x_1},\dots,\lh{x_k})\le\lh{t(\vec
x)}$. Conversely, one can check that $\bas^+$ proves $\lh{xy}\le\lh x+\lh y$; together with other axioms, this
implies that for every term~$t$ (even using $\dotminus$ and~$2^{\min\{u,\lh x\}}$) there is a polynomial $p$ such that
$\bas^+\vdash\lh{t(\vec x)}\le p(\vec{\lh x})$.
\pagebreak[2]
\begin{Lem}\th\label{lem:nf}
Let $\fii(x_1,\dots,x_k)$ be a $\sig0(\dotminus,2^{\min\{u,\lh x\}})$ formula. Then $\bas^+$ proves $\fii(\vec x)$
equivalent to a formula of the form
\begin{multline*}
\LOR_{\sigma_1,\dots,\sigma_k<2^c}\Bigl(\ET_{i=1}^k\bigl(x_i\equiv\sigma_i\pmod{2^c}\bigr)\\
  \land Q_1u_1\le p(\vec{\lh x})\,\cdots\,Q_lu_l\le p(\vec{\lh x})\,
  f_{\vec\sigma}(x_1,\dots,x_k,u_1,\dots,u_l,2^{u_1},\dots,2^{u_l})\ge0\Bigr),
\end{multline*}
where $c$ is a constant, $Q_1,\dots,Q_l\in\{\exists,\forall\}$, $p$ is a nonnegative integer polynomial,
$x_i\equiv\sigma_i\pmod{2^c}$ stands for $x_i=\sigma_i+2^c\lfloor\cdots\lfloor\lfloor x_i/\underbrace{2\rfloor/2\rfloor\cdots/2}_c\rfloor$, and
$f_{\vec\sigma}$ is an integer polynomial.
\end{Lem}
\begin{Pf}
Using the remark before the lemma, we can find a nonnegative integer polynomial~$p$ such that $p(\vec{\lh x})$ bounds
the values of~$\lh t$ for every subterm~$t(\vec x,\vec u)$ occurring in~$\fii$ and all possible values of the
quantified variables~$\vec u$. Then we can rewrite $\fii$ in the form
\[Q_1u_1\le p(\vec{\lh x})\,\cdots\,Q_lu_l\le p(\vec{\lh x})\,\psi(\vec x,\vec u),\]
where $\psi$ is open. The next step is elimination of unwanted function symbols. Let $\lh t$ be a subterm of~$\psi$,
and write $\psi(\vec x,\vec u)=\psi'(\vec x,\vec u,\lh t)$. Then $\psi(\vec x,\vec u)$ is equivalent to
\[\exists u\le p(\vec{\lh x})\,(\lh t=u\land\psi'(\vec x,\vec u,u)).\]
Using the axioms of~$\bas^+$ and the definition of~$2^u$, this is equivalent to
\[\exists u\le p(\vec{\lh x})\,(\half{2^u}\le t<2^u\land\psi'(\vec x,\vec u,u)).\]
Likewise,
\begin{align*}
\psi(\vec x,\vec u,t\#s)&\eq\exists u,v,w\le p(\vec{\lh x})\,
      (u=\lh t\land v=\lh s\land w=uv\land\psi(\vec x,\vec u,2^w)),\\
\psi(\vec x,\vec u,2^{\min\{t,\lh s\}})&\eq\exists u,v\le p(\vec{\lh x})\,
      (u=\lh s\land v=\min\{t,u\}\land\psi(\vec x,\vec u,2^v)),
\end{align*}
where we further eliminate $\lh t$ and $\lh s$ as above, and $\min\{t,u\}$ in an obvious way.
Applying successively these reductions, we can eventually write $\fii$ as
\begin{equation}\label{eq:xxx}
Q_1u_1\le p(\vec{\lh x})\,\cdots\,Q_lu_l\le p(\vec{\lh x})\,\psi(\vec x,\vec u,\vec{2^u}),
\end{equation}
where $\psi$ is an open formula in the language $\p{0,1,+,\cdot,\dotminus,\half x,\le}$.
\begin{Cl}\th\label{cl:terms}
Let $t(\vec x)$ be a $\p{0,1,+,\cdot,\dotminus,\half x}$-term such that the nesting depth of~$\half x$ in~$t$
is~$c$, and the number of occurrences of~$\dotminus$ is~$r$. For every $\vec\sigma<2^c$, there are integer polynomials
$g_1,\dots,g_r$ and $\bigl\{f_{\vec\alpha}:\alpha_1,\dots,\alpha_r\in\{0,1\}\bigr\}$ such that $\bas^+$ proves
\begin{equation}\label{eq:term}
\ET_{i=1}^r(g_i(\vec x)\ge0)^{\alpha_i}\to t(2^c\vec x+\vec\sigma)=f_{\vec\alpha}(\vec x),
\end{equation}
where $\fii^1=\fii$, $\fii^0=\neg\fii$.
\end{Cl}
\begin{Pf*}
By induction on the complexity of~$t$. For example, assume \eqref{eq:term} holds for~$t$, and consider the term $\half
t$. Let $\vec\tau<2$, and assume that $f_{\vec\alpha}(\vec\tau)\equiv\rho\pmod2$, $\rho\in\{0,1\}$. Notice that
all coefficients of $f_{\vec\alpha}(2\vec x+\vec\tau)-\rho$ are even, so $h_{\vec\alpha}(\vec x)=\frac12(f_{\vec\alpha}(2\vec
x)-\rho)$ is again an integer polynomial, and $\bas^+$ proves
\[\ET_{i=1}^r(g_i(2\vec x+\vec\tau)\ge0)^{\alpha_i}\to t(2^{c+1}\vec x+(2^c\vec\tau+\vec\sigma))=\div{f_{\vec\alpha}(2\vec x)}2
=\div{2h_{\vec\alpha}(\vec x)+\rho}2=h_{\vec\alpha}(\vec x).\]
\end{Pf*}
\begin{Cl}\th\label{cl:flas}
Every open formula $\psi(\vec x)$ in the language $\p{0,1,+,\cdot,\dotminus,\half x,\le}$ is equivalent to a
formula of the form
\[\LOR_{\vec\sigma<2^c}\Bigl(\ET_i\bigl(x_i\equiv\sigma_i\pmod{2^c}\bigr)\land\psi_{\vec\sigma}(\vec x)\Bigr)\]
over $\bas^+$, where each $\psi_{\vec\sigma}$ is a Boolean combination of integer polynomial inequalities.
\end{Cl}
\begin{Pf*}
Using \th\ref{cl:terms} and $\bas^+$-provable uniqueness of the representation $x=2^cy+\sigma$, $\sigma<2^c$, we obtain
an equivalent of~$\psi$ in almost the right form except that $\psi_{\vec\sigma}$ is a Boolean combination of
inequalities of the form
\[f(2^{-c}(\vec x-\vec\sigma))\ge0,\]
where $f$ is an integer polynomial. If $d=\deg(f)$, $g(\vec x)=2^{cd}f(2^{-c}(\vec x-\vec\sigma))$ is an integer
polynomial, and the inequality above is equivalent to $g(\vec x)\ge0$.
\end{Pf*}
Let us apply \th\ref{cl:flas} to the formula $\psi(\vec x,\vec u,\vec{2^u})$ in~\eqref{eq:xxx}. Since $\bas^+$ knows
that $2^0=1$ and $2^{u+1}=2\cdot2^u$, we can replace $2^{u_i}\equiv\sigma\pmod{2^c}$ with $2^{u_i}=\sigma\lor(u_i\ge
c\land\sigma=0)$. Moreover, $u_i\equiv\sigma\pmod{2^c}$ can be written as $\exists v\le p(\vec{\lh
x})\,(u_i=2^cv+\sigma)$, and $x_i\equiv\sigma\pmod{2^c}$ can be moved outside the quantifier prefix. Thus, $\fii(\vec
x)$ is equivalent to
\[\LOR_{\vec\sigma<2^c}\Bigl(\ET_{i=1}^k\bigl(x_i\equiv\sigma_i\pmod{2^c}\bigr)
  \land Q_1u_1\le p(\vec{\lh x})\,\cdots\,Q_lu_l\le p(\vec{\lh x})\,\psi_{\vec\sigma}(\vec x,\vec u,\vec{2^u})\Bigr),\]
where $\psi_{\vec\sigma}$ is a Boolean combination of integer polynomial inequalities. We can
reduce~$\psi_{\vec\sigma}$ to a single inequality using
\begin{align*}
\neg(f\ge0)&\eq-f-1\ge0,\\
f\ge0\land g\ge0&\eq\forall v\le p(\vec{\lh x})\,(vf+(1-v)^2g\ge0),
\end{align*}
assuming $p(\vec{\lh x})\ge1$.
\end{Pf}

\begin{Lem}\th\label{lem:seq-intvals}
$\vtcim$ proves the following for every constant~$d$: if $\{f_u:u<n\}$ is a sequence of integer polynomials of
degree~$d$ (each given by a $(d+1)$-tuple of binary integer coefficients), and $a>d$ is a binary integer, there
exists a double sequence $\cmb w=\{w_{u,i}:u<n,i\le d+1\}$ such that $0=w_{u,0}<w_{u,1}<\dots<w_{u,d+1}=a$ and $f_u(x)$ has a
constant sign on each interval $[w_{u,i},w_{u,i+1})$, that is,
\begin{equation}\label{eq:seq}
\forall u<n\,\forall x\ET_{i\le d}\bigl(w_{u,i}\le x<w_{u,i+1}\to(f_u(x)\ge0\eq f_u(w_{u,i})\ge0)\bigr).
\end{equation}
\end{Lem}
\begin{Pf}
Using $\io$, $\{x<a:f(x)\ge0\}$ is a union of at most~$d$ intervals for every polynomial~$f$ of degree at most~$d$,
i.e., $\vtcim$ proves
\[\forall f\,\forall a>d\,\exists 0=x_0<\dots<x_{d+1}=a\,\forall x
   \ET_{i\le d}\bigl(x_i\le x<x_{i+1}\to(f(x)\ge0\eq f(x_i)\ge0)\bigr).\]
Now we would like to invoke $\Sig1$-\acr\ to find a sequence~$w$ satisfying~\eqref{eq:seq},
but we cannot directly do that as the conclusion is only~$\Pii1$.

Let $M\model\vtcim$, and $R$ be its real closure.
Quantifier elimination for RCF furnishes an open formula~$\tet$ in $L_{\M{OR}}$ such that $M\model\tet(x,y,a_0,\dots,a_d)$
iff $f(x)=\sum_{i\le d}a_ix^i$ has no roots in the interval $(x,y]_R$. By replacing $f$ with $2f+1$
if necessary, we may assume $f$ has no integral roots. Let $\alpha_1<\dots<\alpha_c$, $c\le d$, be the list of all
roots of~$f$ in $(0,a]_R$, and let $x_0,\dots,x_{d+1}\in M$ be the sequence of integers
$0,\cl{\alpha_1},\dots,\cl{\alpha_c},a$ (which exist due to $\io$) with duplicates removed, and dummy elements added if
necessary to make it the proper length. Then $f$ has no roots in the intervals $(x_i,x_{i+1}-1]_R$. This means we can
prove in~$\vtcim$ the statement
\[\forall f\,\forall a>d\,\exists 0=x_0<x_1<\dots<x_{d+1}=a\,\ET_{i\le d}\tet(x_i,x_{i+1}-1,f),\]
which has the right complexity, hence we can use $\Sig1$-\acr\ to derive the existence of a sequence~$w$
such that
\[\forall u<n\,\Bigl(w_{u,0}=0\land w_{u,d+1}=a\land
      \ET_{i\le d}\bigl(w_{u,i}<w_{u,i+1}\land\tet(w_{u,i},w_{u,i+1}-1,f_u)\bigr)\Bigr).\]
This implies~\eqref{eq:seq}.
\end{Pf}

\begin{Thm}\th\label{thm:vtc0sig0min}
The $\rsuv$-translation of $\bas^++\sig0(\dotminus,2^{\min\{u,\lh x\}})\text-\Min$, and a fortiori
of~$T^0_2$, is provable in $\vtcim$.
\end{Thm}
\begin{Pf}
Work in $\vtcim$.
It is straightforward but tedious to verify the axioms of~$\bas^+$. Let $\fii(x)$ be (the translation of) a
$\sig0(\dotminus,2^{\min\{u,\lh x\}})$ formula (possibly with other parameters), and $a$ a binary number such that
$\fii(a)$, we have to find the least such number. Since it is enough to do this separately on each residue class
modulo~$2^c$, we can assume using \th\ref{thm:vtc0sig0min} that $\fii(x)$ is equivalent to
\[Q_1u_1\le n\,\cdots\,Q_lu_l\le n\,f(x,\vec u,\vec{2^u})\ge0\]
for $x<a$, where $n$ is a unary number, and $f$ is a polynomial with binary integer coefficients. By \th\ref{lem:seq-intvals}, there is a sequence~$w$ such that
\[w_{\vec u,i}\le x<w_{\vec u,i+1}\to(f(x,\vec u,\vec{2^u})\ge0\eq f(w_{\vec u,i},\vec u,\vec{2^u})\ge0)\]
for all $x<a$, $\vec u\le n$, and~$i\le d$. As $\vtc$ proves that every sequence of integers can be sorted, there is an increasing
sequence 
$\{w'_j:j<m\}$ whose elements include every $w_{\vec u,i}$. Consequently, the truth value of $\fii(x)$ is constant
on each interval $[w'_j,w'_{j+1})$, and the minimal $x<a$ satisfying $\fii(x)$, if any, is $w'_{j_0}$, where
\[j_0=\min\{j<m:\fii(w'_j)\}.\]
The latter exists by $\Sig0(L_{\ob\vtcim})\text-\comp$.
\end{Pf}

We remark that the proof used nothing particularly special about division by~$2$, except that $\bas$ conveniently includes
the $\half x$ function symbol and the relevant axioms. We could allow more general instances of division as
long as the values of all denominators encountered when evaluating a $\sig0$ formula on $[0,a]$ have a common multiple
which is a length (unary number); in particular, \th\ref{thm:vtc0sig0min} (along with an appropriate version of \th\ref{lem:nf}) holds for $\sig0$
formulas in a language further expanded by function symbols for $\tdive x{\lh{\lh y}}$ and $\fl{x/\max\{1,\lh{\lh y}\}}$.

We formulated \th\ref{thm:vtc0sig0min} for $\vtcim$ as we have been working with this two-sorted theory throughout the main part of
the paper, however here it is perhaps more natural to state the result directly in terms of one-sorted arithmetic to avoid
needless $\rsuv$ translation. A theory $\delt1\text-\M{CR}$ corresponding to~$\tc$ was defined by Johannsen and Pollett
\cite{joh-pol:d1cr}, and shown $\rsuv$-isomorphic to $\vtc$ by Nguyen and Cook \cite{ngu-cook}. Recall also Johannsen's
theory $C^0_2[\M{div}]$ from Section~\ref{sec:imul-div}.
\begin{Cor}\th\label{cor:joh}
The theories $\delt1\text-\M{CR}+\imul$ and $C^0_2[\M{div}]$ prove $\sig0(\dotminus,2^{\min\{u,\lh
x\}})\text-\Min$ and therefore $T^0_2$.
\noproof\end{Cor}

To put \th\ref{thm:vtc0sig0min} in context, there has been a series of results to the effect that various subsystems of
bounded arithmetic axiomatized by sharply bounded schemata are pathologically weak. Takeuti~\cite{tak:s02} has shown
that $S^0_2=\sig0\text-\pind$ does not prove the totality of the predecessor function, and Johannsen~\cite{joh:s02}
extended his method to show that $S^0_2$ in a language including $\dotminus$, $\tdive xy$, and bit counting does not
prove the totality of division by three (or even of the $\Ac$ function $\fl{2^{\lh x}/3}$). Boughattas and
Ko\l odziejczyk~\cite{bou-kol}  have shown that $T^0_2=\sig0\text-\ind$ does not prove that nontrivial divisors of
powers of two are even, and by Ko\l odziejczyk~\cite{kol:t02}, it does not even prove $3\nmid2^{\lh x}$. These results
also apply to certain mild extensions of~$T^0_2$, nevertheless no unconditional independence result is known for
$\sig0$-\Min, or its subtheory $T^0_2+S^0_2$.

What makes such separations possible is a lack of computational power. It is no coincidence that there are no result of
this kind for two-sorted Zambella-style theories, where already the base theory~$V^0$ proves the totality of all
$\Ac$-functions: we can show $V^0(p)\nsset V^0(q)$ for primes $p\ne q$ using the known lower bounds for~$\Ac[p]$, but we
have no independence results for stronger theories without complexity assumptions such as $\Ac[6]\ne\ph$. This is
directly related to the expressive power of sharply bounded formulas: while $\Sig0$ formulas can define all $\Ac$
predicates, the ostensibly quite similar $\sig0$ formulas (that even involve the $\tc$-complete multiplication
function) have structural properties that preclude this, as witnessed by Mantzivis's result. Indeed, the pathological
behaviour of~$T^0_2$ disappears if we slightly extend its language: as proved in~\cite{ej:t02}, $T^0_2(\tdive
xy)=\tpv$, and this can be easily extended to show $\sig0(\tdive xy)\text-\Min=T^1_2$.

\th\ref{thm:vtc0sig0min} formally implies only conditional separations: in particular, $\tpv\nsset\sig0\text-\Min$
unless $\ptime=\tc$, and $\sig0\text-\Min\ssset T^1_2$ unless $\ph=\cxt{BH}\sset\tc/\poly$ and $\cxt{PLS}=\cxt{FTC}^0$
(provably in $\sig0$-\Min). However, heuristically it gives us more. If $\sig0$-\Min\ were a ``computationally
reasonable'' theory, we would expect it to coincide with~$T^1_2$ due to its shape, or at the very least to correspond to a
class closer to~$\cxt{PLS}$ than~$\tc$. Thus, \th\ref{thm:vtc0sig0min} indicates that it might be a pathologically
weak theory in some way, and therefore amenable to unconditional independence results by means of a direct
combinatorial construction of models in the spirit of~\cite{bou-kol,kol:t02}.

\section{Conclusion}\label{sec:conclusion}
The weakest theory of bounded arithmetic in the setup of \cite{zamb:notes,cook-ngu} that can talk about elementary
arithmetic operations on binary integers is~$\vtc$. We have shown that its strengthening $\vtcim$ proves that these
operations are fairly well behaved in that they satisfy open induction. Despite that the theory $\vtcim$ corresponds to the
complexity class~$\tc$ similarly to~$\vtc$, it is still an interesting problem what properties of integer arithmetic
operations are provable in plain~$\vtc$. In view of \th\ref{thm:vtciopen,cor:vtc-div-imul}, we have:
\begin{Cor}\th\label{cor:vtc-div-iop}
$\vtc$ proves~$\io$ if and only if it proves~$\Div$.
\noproof\end{Cor}
\begin{Que}\th\label{q:vtc-div}
Does $\vtc$ prove $\Div$? In particular, does it prove the soundness of the division algorithm by Hesse
et~al.~\cite{hab}?
\end{Que}
While the analysis of the algorithm in~\cite{hab} generally relies on quite elementary tools, its
formalization in~$\vtc$ suffers from ``chicken-and-egg'' problems. For instance, the proof of Lemma 6.1, whose goal is
to devise an algorithm for finding small powers in groups, assumes there is a well-behaved powering function, and uses
its various properties to establish that its value is correctly computed by the algorithm. This is no good if we need
the very algorithm to construct the powering function in the first place. Similarly, integer division is employed
throughout Section~4. It is not clear whether one can circumvent these circular dependencies in~$\vtc$. On the other
hand, the requisite operations such as division are available in $\vtcim$, which makes it plausible that $\vtcim$
\emph{can} formalize the arguments.

We remark that it is not difficult to do division by \emph{standard} integers in~$\vtc$. This means $\vtc$ knows that
binary integers form a \emph{$\Z$-ring}, and in particular, they satisfy all universal consequences of~$\io$ by a
result of Wilkie~\cite{wil:iop}. ($\io$ itself is a $\forall\exists$-axiomatized theory, and likewise, $\Div$ is a
$\forall\exists$ sentence.)

As explained in Section~\ref{sec:buss-th}, our main result implies that $\vtcim$ (or better, the corresponding
one-sorted theory $\delt1\text-\M{CR}+\imul$) proves minimization for $\sig0$ formulas in Buss's language, which
suggests that the theory axiomatized by $\sig0$-\Min\ is rather weak. Consequently, it might be feasible to
unconditionally separate this theory from stronger fragments of~$S_2$, nevertheless our argument gives no clue how to
do that.
\begin{Prob}\th\label{prob:sep}
Prove that $\sig0$-\Min\ is strictly weaker than~$T^1_2$ without complexity-theoretic assumptions.
\end{Prob}

\subsection*{Acknowledgement}
I am indebted to Leszek Ko\l odziejczyk for useful suggestions on an earlier version of this paper.

\bibliographystyle{mybib}
\bibliography{mybib}
\end{document}

%% file: defs.tex
\ifx\mydefs\undefined\else \fi
\let\mydefs\relax

\allowdisplaybreaks[2]

\ifx\loadcyr\undefined\else
\input cyracc.def
\makeatletter
 at 1\@ptsize pt
 at 1\@ptsize pt
 at 1\@ptsize pt
\makeatother

\fi

\let\M\mathit
\def\gobble#1{}
\def\fixsup#1#2{{#1\let\dp\gobble\mathstrut}^#2_}
\let\mr\mathrel

\def\bme{\hskip.75em\relax}

\let\eq\leftrightarrow
\let\EQ\Leftrightarrow

\def\iff{\quad\text{iff}\quad}
\let\LOR\bigvee
\let\ET\bigwedge

\def\?{\mathbin?}

\let\model\vDash

\newbox\circlebox
\setbox\circlebox\hbox{$\bigcirc$}
\def\circled#1{%
  \setbox0\hbox to\wd\circlebox{\hss$#1$\hss}\wd0=0pt
  \box0\copy\circlebox}

\let\fii\varphi
\let\tet\vartheta
\let\ep\varepsilon

\def\greek#1{$\expandafter\greeknum\csname c@#1\endcsname$}
\makeatletter
\def\greeknum#1{\ifcase#1\or\alpha\or\beta\or\gamma\or\delta\or\ep
      \or\digamma\or\zeta\or\eta\or\tet\or\iota\else\@ctrerr\fi}
\makeatother

\def\p#1{\langle#1\rangle}

\def\lh#1{\lvert#1\rvert}
\let\abs\lh
\def\Abs#1{\left|#1\right|}

\let\sset\subseteq
\let\nsset\nsubseteq
\let\ssset\subsetneq
\let\Sset\supseteq

\let\onto\twoheadrightarrow

\newcommand\rpair[3][3em]{\mathrel{%
   \begin{matrix}%
     \strut\smash{\xrightonto{\hbox to#1{\hss$#2$\hss}}}\\[-1.7ex]%
     \strut\smash{\xleftembed[\hbox to#1{\hss$#3$\hss}]{}}%
   \end{matrix}}}
\makeatletter
\newcommand\xrightonto[2][]{\ext@arrow 0359\rightontofill{#1}{#2}}
\newcommand\xleftembed[2][]{\ext@arrow 3095\leftembedfill{#1}{#2}}
\def\leftembedfill{\arrowfill@\leftarrow\relbar\hookleftnoarrow}
\def\rightontofill{\arrowfill@\relbar\relbar\onto}
\def\hookleftnoarrow{\DOTSB\relbar\joinrel\rhook}
\makeatother

\def\fl#1{\lfloor#1\rfloor}

\def\cl#1{\lceil#1\rceil}

\def\div{\genfrac\lfloor\rfloor{}{}}

\def\tdive#1#2{\fl{#1/2^{#2}}}
\def\half#1{\fl{#1/2}}

\mathchardef\#="2023 

\def\rsuv{\M{RSUV}}

\def\tsum{\textstyle\sum}
\providecommand\dotminus{\mathbin{\mathchoice
       {\scriptstyle\dodotminus\displaystyle}%
       {\scriptstyle\dodotminus\textstyle}%
       {\scriptscriptstyle\dodotminus\scriptstyle}%
       {\scriptscriptstyle\dodotminus\scriptscriptstyle}}}
\def\dodotminus#1{\dot{\smash{#1-}}}

\ifx\busspf\undefined\else
\usepackage{bussproofs}
\EnableBpAbbreviations

\fi

\ifx\symlasy\undefined

  \def\centerdot#1{{%
    \setbox0\hbox{$\mathop{#1}$}\dimen0 \ht0
    \setbox0\hbox{$#1$}\advance\dimen0 -\ht0
    \setbox2\hbox to\wd0{\hss$\mathop{\cdot}$\hss}\wd2=0pt
    \lower\dimen0\box2\box0 }}
\else
  
  \def\centerdot#1{{%
     \setbox0\hbox{$#1$}%
     \raise0.206\ht0\hbox to\wd0{\hss$\cdot$\hss}%
     \kern-\wd0 \box0 }}
\fi

\let\sls|

\def\Up{{\setbox0\hbox{$\uparrow$}%
         \lower\dp0\hbox to\wd0{\hss\vrule width4pt height.4pt\hss}%
         \kern-\wd0\box0}}
\def\UP{{\setbox0\hbox{$\uparrow$}%
         \lower\dp0\hbox to\wd0{\hss\vrule width4pt height.4pt\hss}%
         \kern-\wd0\copy0\kern-\wd0\raise.35ex\box0}}
\def\Down{{\setbox0\hbox{$\downarrow$}%
         \raise\ht0\hbox to\wd0{\hss\vrule width4pt depth.4pt\hss}%
         \kern-\wd0\box0}}

\newif\ifnadm

\def\doadm{\mathrel{%
   \setbox0 \hbox{$\mathop\vdash$}\dimen0 \ht0
   \setbox0 \hbox{$\vdash$}\advance\dimen0 -\ht0
   \vrule width.8\fontdimen8 \textfont3 height\ht0 depth\dp0
   \mkern-1mu
   \lower\dimen0 \hbox{$\vcenter{%
      \ifnadm
        \setbox0 \hbox{$\scriptstyle\sim\mathstrut$}%
        \hbox{\hbox to\wd0{\hss$\scriptstyle/$\hss}\kern-\wd0 \box0 }%
      \else
        \hbox{$\scriptstyle\sim\mathstrut$}%
      \fi}$}}}

\def\nrstyle#1#2#3{%
  \setbox0\hbox{$#1\bigcirc$}%
  \vcenter{\hbox to\wd0{\hss$#2#3$\hss}}%
  \kern-\wd0\box0 }

\DeclareMathOperator\card{card}

\DeclareMathOperator\rcl{rcl}

\let\cxt\mathrm

\def\ptime{\cxt P}

\def\ph{\cxt{PH}}

\DeclareMathOperator\poly{poly}

\def\tc{\cxt{TC}^0}
\def\Ac{\cxt{AC}^0}
\def\LT{\cxt{DLOGTIME}}
\def\task#1{{\normalfont\textsc{#1}}}

\def\sig{\Sigma^b_}

\def\delt{\Delta^b_}
\def\Sig{\Sigma^B_}
\def\Pii{\Pi^B_}

\def\st{\expandafter\hat}

\def\io{\M{IOpen}}
\def\idz{I\Delta_0}
\def\bas{\M{BASIC}}

\def\tpv{PV_1}

\def\vtc{\M{VTC}^0}
\def\Div{\M{DIV}}
\def\imul{\M{IMUL}}
\def\vtcim{\vtc+\imul}

\def\BB{\M{BB}}

\def\sch#1{\ensuremath{\M{#1}}}

\def\Min{\sch{MIN}}
\def\lind{\sch{LIND}}
\def\ind{\sch{IND}}
\def\pind{\sch{PIND}}

\def\comp{\sch{COMP}}

\def\ac{\sch{AC}}
\def\acr{\sch{AC^R}}

\def\Q{\mathbb Q}
\def\Z{\mathbb Z}
\def\RR{\mathbb R}

\mathcode`\*="0203
\let\ob\overline

\mathchardef\mhyphen="2D

\def\noproof{\leavevmode\unskip\bme\vadjust{}\nobreak\hfill$\qed$\par}
\let\qed\Box
\newenvironment{Pf}[1][]
  {\par\noindent\textit{Proof\optpar{#1}:}\bme\ignorespaces}
  {\noproof\pagebreak[2]\vskip\medskipamount\ignorespacesafterend}
\def\optpar#1{\ifx\relax#1\relax\else\ #1\fi}
\def\qedhere{\relax\ifmmode\eqno\qed\expandafter\aftergroup
                   \else\noproof\fi\noqed}
\def\noqed{\let\noproof\relax}

\theoremstyle{plain}
\ifx\shortthm\undefined
\newtheorem{Thm}{Theorem}[section]
\else
\newtheorem{Thm}{Theorem}
\fi
\newtheorem{Prop}[Thm]{Proposition}
\newtheorem{Cor}[Thm]{Corollary}
\newtheorem{Lem}[Thm]{Lemma}

\newtheorem{Que}[Thm]{Question}
\newtheorem{Prob}[Thm]{Problem}

\newtheorem{Cl}{Claim}[Thm]
\ifx\shortthm\undefined
\def\theCl{\arabic{Cl}}
\fi

\theorembodyfont\upshape
\newtheorem{Def}[Thm]{Definition}
\newtheorem{Rem}[Thm]{Remark}
\newtheorem{Exm}[Thm]{Example}

\newenvironment{Pf*}{\let\qed\qedCl\Pf}\endPf

\def\theequation{\arabic{equation}}
\def\numberthis{\addtocounter{equation}{1}\tag{\theequation}}

\usepackage[reftex]{theoremref}

\newif\iflinenumbers
\linenumberstrue

{\catcode`\^^I=13 \catcode`\^^M=13
\gdef\doalgo#1#2\end#{\hbox to\hsize{\hss \let^^I\qquad%
  \def\\^^M{\nobreak\hfil\break\vadjust{}\qquad}%
  \fboxsep1em \linenum0 %
  \fbox{\hsize#1\vbox{%
  \everypar{\advance\linenum1 %
      \hbox to1.2em{%
           \hss\iflinenumbers$\scriptstyle\the\linenum$\hskip.6em\fi}}%
  #2}}\hss}\end}}
\newcount\linenum

\def\key{\relax\ifmmode\expandafter\mathbf\else\expandafter\textbf\fi}

\def\allowhyphens{\nobreak\hskip0pt\relax}

\DeclareRobustCommand*\magiclparen{\ifmmode(\else\textup(\allowhyphens\fi}
\DeclareRobustCommand*\magicrparen{\ifmmode)\else\textup)\fi}
\let\lparen=(  \let\rparen=)
\def\magicparon{\catcode`\(\active\catcode`\)\active}
\def\magicparoff{\catcode`\(12 \catcode`\)12 }
\AtBeginDocument{\ifx\ifPreview\iftrue\else\magicparon\fi}
\magicparon
\let (=\magiclparen  \let )=\magicrparen

\ifx\sameenum\undefined
  
  \ifx\enumup\undefined
    
  \else
    
  \fi

\else

\fi

\magicparoff

\def\cmb{\mathcode`\,"8000 }
\mathchardef\comma=\mathcode`\,
{\catcode`\,=\active \gdef,{\comma\penalty\relpenalty}}

\providecommand\dedic{\thanks{Supported by
grant IAA100190902 of GA AV \v CR, Center of Excellence CE-ITI under the grant
P202/12/G061 of GA \v CR, and RVO: 67985840. The research leading to these
results has received funding from the
European Research Council under the European Union's Seventh Framework
Programme (FP7/2007--2013)~/ ERC grant agreement no.~339691.}}
\author{Emil Je\v r\'abek\dedic\\[\medskipamount]
Institute of Mathematics of the Academy of Sciences\\
\small \v Zitn\'a 25,
115\:67 Praha 1,
Czech Republic,
email: \texttt{jerabek@math.cas.cz}
}